\documentclass[12pt]{iopart}
\usepackage[utf8]{inputenc}
\usepackage[pdftex]{graphicx}
\usepackage[colorlinks=true,citecolor=blue]{hyperref}
\usepackage{braket}

\usepackage{cite}
\usepackage{bbm}
\usepackage{iopams}
\expandafter\let\csname equation*\endcsname\relax
\expandafter\let\csname endequation*\endcsname\relax
 \usepackage{amsmath}

\newcommand{\onlinecite}{\cite}

\newcommand{\tx}{\text}

\renewcommand{\vec}[1]{\ensuremath{\mathbf{#1}}}

\newcommand{\Gm}{\Gamma}
\newcommand{\sg}{\sigma}

\newcommand{\Abs}[1]{\ensuremath{\left| #1 \right|}}

\begin{document}
\title{Inverse counting statistics based on generalized factorial cumulants}

\author{Philipp Stegmann and J\"urgen K\"onig}

\address{Theoretische Physik, Universit\"at Duisburg-Essen and CENIDE, D-47048 Duisburg, Germany}
\ead{philipp.stegmann@uni-due.de}

\date{\today}

\begin{abstract}
We propose a procedure to reconstruct characteristic features of an unknown stochastic system from the long-time full counting statistics of some of the system's transitions that are monitored by a detector.
The full counting statistics is conveniently parametrized by so-called generalized factorial cumulants.
Taking only a few of them as input information is sufficient to reconstruct important features such as the lower bound of the system dimension and the full spectrum of relaxation rates.
The use of generalized factorial cumulants reveals system dimensions and rates that are hidden for ordinary cumulants. 
We illustrate the inverse counting-statistics procedure for two model systems: a single-level quantum dot in a Zeeman field and a single-electron box subjected to sequential and Andreev tunneling.
\\\\\\
{\it Keywords\/}: full counting statistics, stochastic systems, factorial cumulants,\\
\hphantom{{\it Keywords\/}: }Coulomb blockade, single-electron tunneling, Andreev tunneling
\end{abstract}

\submitto{\NJP}

\maketitle

\section{Introduction}
A stochastic system is characterized by the rates of the stochastic transitions between the possible states.
The dynamics of the stochastic system may be probed by a detector that is sensitive to one or several (but, in general, not all) of the possible transitions.
In biological physics, stochastic transitions like the steps of motor proteins~\cite{kolomeisky_molecular_2007, chemla_exact_2008}, intramolecular conformational changes~\cite{zhou_detecting_2011, choi_single_2012, chung_single_2012}, and enzymatic turnovers generating fluorescent products~\cite{english_ever_2006, moffitt_extracting_2014} have been studied. Detectors are optical tweezers~\cite{kolomeisky_molecular_2007, cornish_survey_2007}, atomic force, or fluorescence microscopes~\cite{cornish_survey_2007, kim_single_2013}.
In mesoscopic physics, discrete charge-transfer events~\cite{levitov_electron_1996, bagrets_full_2003} through single-~\cite{lindebaum_spin-induced_2009, schmidt_charge_2009, schaller_counting_2010} and multi-level~\cite{belzig_full_2005} quantum dots, interferometers~\cite{urban_coulomb-interaction_2008}, superconducting~\cite{belzig_full_counting_2001, boerlin_full_2002, cuevas_full_2003, johansson_full_2003, pilgram_noise_2005, morten_full_2008, braggio_superconducting_2011} and feedback-controlled systems~\cite{poeltl_feedback_2011, daryanoosh_stochastic_2016, wagner_squeezing_2016} have been studied.
Charge transfers can be detected by a quantum point contact~\cite{gustavsson_counting_2005, fujisawa_bidirectional_2006, gustavsson_measurements_2007, fricke_bimodal_2007, flindt_universal_2009, gustavsson_electron_2009, fricke_high_2010, fricke_high-order_2010, komijani_counting_2013}, a single-electron transistor~\cite{martinis_metrological_1994, dresselhaus_measurement_1994, lotkhov_storage_1999, lu_real_2003, bylander_current_2005}, optical~\cite{kurzmann_optical_2016}, or interferometric detectors~\cite{dasenbrook_dynamical_2016}.

Counting the number $N$ of transitions within a time interval $[0;t]$ repeatedly many times yields the probability distribution $P_N(t)$, referred to as {\it full counting statistics}.
For a well-characterized system, the states, the rates between them, and the coupling to the detector are known.
It is, then, straightforward to compute the full counting statistics and compare it with experimentally measured data.
Suppose, however, that the underlying model for a stochastic system is unclear and the only information available is the counting statistics measured by the detector.
It is, then, desirable to have a systematic approach to distill out of the measured counting statistics the relevant information for reconstructing properties of the underlying model, including basic properties such as the number of states of the stochastic system.
Such an approach can be dubbed {\it inverse counting statistics}~\cite{bruderer_inverse_2014}.

What are the properties of the stochastic system that one may hope to reconstruct by inverse counting statistics?
First of all, there is the number $M$ of possible system states. 
Second, the stochastic system is characterized by the spectrum of relaxation rates with which it relaxes back to its (equilibrium or nonequilibrium) steady state after being disturbed externally~\cite{splettstoesser_charge_2010, pulido_time_2012, schulenborg_detection_2014, schulenborg_parity_2016, riwar_readout_2016, vanherck_relaxation_2016}.
Of course, inverse counting statistics cannot distinguish between different stochastic systems that are equivalent in the sense that they produce the same counting statistics, even if the stochastic systems possess a different numbers of states.
Therefore, inverse counting statistics can at most deliver the {\it minimal} number of system states necessary that is compatible with the observed counting statistics.

To make the inverse counting statistics a powerful and practical tool, one should keep the circumstances for the acquisition of the input data as transparent and as simple as possible.
Therefore, we restrict ourselves to the following scenario.
First, only steady-state counting statistics is considered, i.e., we assume that the system has already relaxed before counting starts.
This excludes studying transient behavior after a perturbation of the system.
The latter would, on the one hand, offer a direct access to some relaxation rate of the system~\cite{feve_on-demand_2007, mahe_current_2010, beckel_asymmetry_2014, hofmann_measuring_2016,kurzmann_electron_2017}.
On the other hand, the determination of the full spectrum of relaxation rates would require the knowledge of how to perturb the system in order to probe a specific relaxation rate.

Second, we concentrate on the limit of long measuring-time intervals $[0;t]$, for which the system dynamics is dominated by the slowest relaxation rate only. 
Nevertheless, the procedure of inverse counting statistics yields the full spectrum of all relaxation rates, as explained below.

Given a measured distribution $P_N(t)$, what are the input data for the inverse counting statistics?
In~\onlinecite{bruderer_inverse_2014}, it was suggested to use the (long-time) cumulants of the distribution function.
Here, we propose to use generalized factorial cumulants~\cite{stegmann_detection_2015, stegmann_short_2016} instead.
The advantage of the latter is that they depend on an arbitrarily chosen parameter $s$. The outcome of the inverse counting statistics (such as the number of system states or the spectrum of relaxation rates) should, however, not depend on this parameter~$s$.
Therefore, the $s$-independence of the results defines a powerful consistency criterion. Furthermore, as we will see in section~\ref{sec:SEBAndreev}, there are special cases in which part of the relaxation-rate spectrum is not accessible by inverse counting statistics with ordinary cumulants but is detectable by using generalized factorial cumulants with properly chosen parameter~$s$. 

As another difference to~\onlinecite{bruderer_inverse_2014}, we allow for a more general system-detector coupling by introducing the {\it counting power} $m$. In~\onlinecite{bruderer_inverse_2014}, the detector is assumed to be sensitive to only a single transition between two specific states increasing the detector counter just by one, the counting power is $m=1$. If this transition increases the detector counter by $k$, the counting power is $m=k$. 
If several transitions are counted by the detector, the counting power can be even larger. We allow for detectors counting arbitrarily many transitions between arbitrarily many states increasing the detector counter by an arbitrary amount.   
Therefore, our inverse counting procedure does not only test compatibility with the number $M$ of system states but also with the counting power $m$.

The paper is organized as follows. In section~\ref{sec:FCS}, we give a short introduction how full counting statistics, especially generalized factorial cumulants, are calculated by means of a Markovian master equation. Subsequently, in section~\ref{sec:ics}, we explain the general procedure of inverse counting statistics. This procedure is, then, illustrated in sections~\ref{sec:QDZee} and \ref{sec:SEBAndreev} for two model systems: a single-level quantum dot in a Zeeman field and a single-electron box subjected to sequential and Andreev tunneling. In section~\ref{sec:summary}, we give a short summary of the inverse counting procedure introduced in this paper.

\section{Full counting statistics for stochastic systems}\label{sec:FCS}
\begin{figure}[t]
\begin{center}
\includegraphics[scale=1.00]{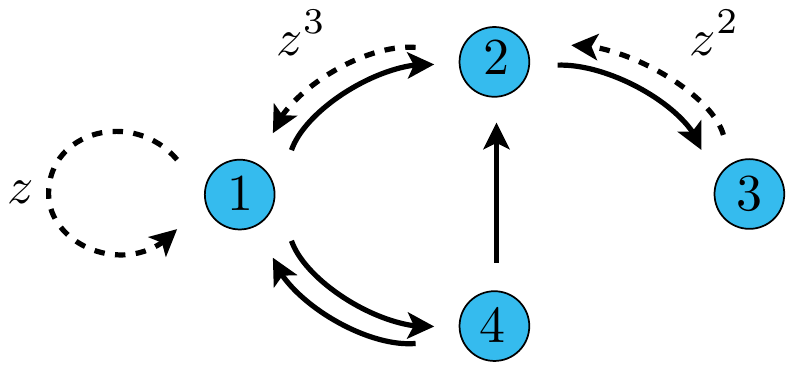}
\caption{
	Stochastic system with $M=4$ states. Transitions are indicated by arrows. Dashed arrows indicate transitions counted by the detector with counting factors~$z^k$.
		}
\label{fig1}
\end{center}
\end{figure}

In this paper, we consider stochastic systems as represented in figure~\ref{fig1}, i.e., systems whose time evolution is completely governed by a Markovian master equation.
There is a finite number $M$ of system states, labeled by $\chi$. Systems with coherent superposition of different $\chi$ are not taken into account. Arrows indicate transitions from state~$\chi$ to state~$\chi'$ with a transition rate $\Gamma_{\chi' \chi}$. Non-Markovian effects~\cite{braggio_full_2006, flindt_counting_2008, flindt_counting_2010} are not taken into account. The states of the master equation can be, e.g., three charge states of a metallic island coupled to one superconducting lead as discussed in the section~\ref{sec:SEBAndreev}, while the continuum of electronic states on the metallic island and lead enters only effectively via the transition rates.

The counting factors $z^k$ with $k=1,2,\ldots$ next to the dashed arrows specify the coupling of the detector to the system: if the system undergoes such a transition, the number $N$ counted by the detector increases by $k$. There may be also processes with rate $\Gamma_{\chi\chi}$ monitored by the detector in which the system ends up in the same state $\chi$ as it started, which is, e.g., the case for cotunneling through a magnetic atom~\cite{sothmann_nonequilibrium_2010}.

Without the detector, the master equation for the probabilities $p^{\chi}(t)$ to find the system in state $\chi$ takes the form
\begin{equation}
\label{eq:master}
	\dot p^\chi(t)= \sum_{\chi'} [\Gamma_{\chi \chi'} p^{\chi'}(t) - \Gamma_{\chi' \chi} p^{\chi}(t)] \, .
\end{equation}
The first term of the sum describes transitions into state $\chi$, the second term transitions out of state $\chi$.
To account for the detector, we need to replace this master equation for $p^\chi(t)$ by an $N$-resolved one for $p_N^\chi(t)$, where $N$ is the number of detector counts in the time interval $[0;t]$.
For this, we introduce the coupling constants $d_{\chi\chi'}^k$ which is $1$ if the detector count is increased by $k$ for the transition from $\chi'$ to $\chi$, and $0$ otherwise.
We obtain
\begin{equation}\label{eq:masterNresolved}
	\dot p^\chi_N(t)= \sum_{\chi'}\sum_k  d_{\chi\chi'}^k \Gamma_{\chi \chi'} p^{\chi'}_{N-k}(t) - \sum_{\chi'} \Gamma_{\chi' \chi} p^{\chi}_{N}(t)
\, .
\end{equation}
Solving for the $N$-resolved probabilities yields the counting statistics of the detector via $P_N(t)=\sum_{\chi}p_N^\chi(t)$.

We now turn to the question of how to extract from a given distribution $P_N(t)$ the information that serves as an input for the inverse counting statistics.
One possibility would be to use the {\it ordinary} (long-time) cumulants of the distribution function, as suggested in~\onlinecite{bruderer_inverse_2014}.
The cumulants are obtained as derivatives $C_k(t)=\partial_z^k \ln {\cal M}(z,t)| _{z=0}$ of the generating function ${\cal M} (z,t)=\sum_{N=0}^\infty e^{N z} P_N(t)$.
Here, we propose, as an alternative, to employ generalized factorial cumulants~\cite{stegmann_detection_2015} in the long-time limit.
Generalized factorial cumulants are derived from the generating function 
\begin{equation}
\label{generating_function}
	{\cal M}_s(z,t):=\sum\limits_{N=0}^\infty (z+s)^N P_N(t) \, ,
\end{equation}
by evaluating the derivatives
\begin{equation}
\label{eq:Csk}
	C_{s,k}(t):=\frac{\partial^k {\ln \cal M}_s(z,t)}{\partial z^k} \bigg|_{z=0} = \frac{\partial^k {\ln \cal M}_0(z,t)}{\partial z^k} \bigg|_{z=s}\, .
\end{equation}
Counting factors enter the generating function~(\ref{generating_function}) as powers of $z$ and not $e^z$ as in the case of ordinary cumulants, which simplifies analytic expressions in the following.

The special case $s=1$ recovers the factorial cumulants recently introduced in the context of mesoscopic transport~\cite{kambly_factorial_2011, kambly_time-dependent_2013}.
They are called {\it factorial} because the corresponding moments $\partial^k_z {\cal M}_1(z,t)|_{z=0} =\langle N^{(k)} \rangle$ are expectation values of the {\it factorial} power $N^{(k)}:= N (N-1) \ldots (N-k+1)$ instead of the ordinary power $N^k$ which defines ordinary moments.
{\it Generalized} factorial cumulants depend on an extra parameter~$s$ that describes a shift of the complex variable $z$ by the amount $s-1$ along the real axis in the complex plane. In contrast to previous works~\cite{stegmann_detection_2015, stegmann_short_2016}, we define the generating function without a normalization factor $1/{\cal M}_s(0,t)$.
Such a $z$-independent normalization factor would not influence the cumulants of order $k>0$, but it would set, per definition, $C_{s,0}=0$.
Without this normalization factor, $C_{s,0}(t)=\ln \sum_Ns^NP_N(t)$ contains non-trivial information that can be used for the inverse counting statistics.
Both factorial ($s=1$) and generalized factorial cumulants ($s\neq 1$) have been utilized to detect correlations in charge-transfer statistics~\cite{kambly_factorial_2011, kambly_time-dependent_2013, stegmann_detection_2015, droste_finite_2016, stegmann_short_2016}.

In the context of inverse counting statistics, the parameter~$s$ will play a very important role in two respects.
First, since the underlying stochastic model for a measured distribution $P_N(t)$ is independent of $s$, the outcome of the inverse counting statistics must also be independent of the parameter~$s$.
Therefore, the required $s$-independence of the obtained results defines a criterion for the compatibility of the measured data with the assumed underlying model. 
Second, as we will see below, there are special cases in which the inverse counting statistics with factorial cumulants ($s=1$) would indicate compatibility with a too small stochastic system and only generalized factorial cumulants ($s\neq 1$) reveal the higher dimension of the underlying stochastic model.

To calculate the generalized factorial cumulants for a given stochastic system, it is convenient to first perform a $z$-transform of the $N$-resolved master equation~(\ref{eq:masterNresolved}), i.e., multiply with $z^N$ and then sum over $N$. If we combine the $z$-transformed $N$-resolved probabilities $p_z^\chi = \sum_N z^N p_N^{\chi}$ of the different states $\chi$ in a vector ${\bf p}_z(t)$, the $z$-transformed master equation can be written in the form 
\begin{equation}\label{eq:masterZ}
	\dot{\vec{p}}_{z}(t)= \vec{W}_{z} \vec{p}_{z}(t) \, ,
\end{equation} 
with matrix elements
\begin{equation}
	\left(\vec{W}_{z} \right)_{\chi \chi'} =
	\sum_k z^k  d_{\chi\chi'}^k \Gamma_{\chi \chi'} - \delta_{\chi\chi'} \sum_{\chi''} \Gamma_{\chi''\chi'} \, .
\end{equation}
Two examples for $\vec{W}_z$ are given in equations~\eqref{eq:Wzexampledot} and \eqref{eq:Wzexample}. 
For $z=1$, equation~(\ref{eq:masterZ}) is nothing but the master equation~(\ref{eq:master}).

The solution of equation~\eqref{eq:masterZ} is $\vec{p}_{z}(t) = \exp \left( \vec{W}_{z}t \right) \vec{p}_z(0)$. 
Since $p_N^\chi(0)\sim \delta_{N,0}$, the initial vector $\vec{p}_z(0)$ is independent of $z$ and describes the initial probability distribution. 
The matrix exponential $ \exp \left( \vec{W}_{z}t \right)=\sum_{j=1}^M \exp \left[\lambda_j(z) t \right]\vec{r}_{j,z} \otimes \vec{l}_{j,z}^{T}$ can be expressed in terms of the eigenvalue spectrum $\{\lambda_j(z) \}$ of $\vec{W}_z$ by making use of the decomposition into the left and right eigenvector $\vec l_{j,z}$ and $\vec r_{j,z}$ with normalization $\vec{l}_{j,z}^{T}\cdot \vec{r}_{j',z}=\delta_{j j'}$. 
For an arbitrary initial distribution $\vec{p}_z(0)$ of the system, the systems relaxes exponentially in time to its steady state, governed by the eigenvalues $\lambda_j(z)$. The eigenvalues $\lambda_j(z)$ at $z=1$ are the system's relaxation rates mentioned in the introduction. They are either real or they appear as complex-conjugated pairs.
In the following, we assume that counting starts only after the system has reached its steady state, i.e., $\vec{p}_z(0)$ is the stationary probability distribution, determined by $\vec{W}_1 \vec{p}_z(0)=0$ and $\vec{e}^T\cdot \vec{p}_z(0)=1$, where we defined $\vec{e}^T=(1,\ldots,1)$ to sum over all states $\chi$ in $\vec{p}_z(0)$.

Finally, taking into account that the generating function can be written as ${\cal M}_s(z,t) =  \vec{e}^T \cdot \vec{p}_{z+s}(t)$, we obtain  
\begin{equation}
	\label{eq:decompcum}
	C_{s,k}(t) =\frac{\partial^k}{\partial z^k} \ln \left[\sum_{j=1}^M \left( \vec{e}^T \cdot \vec{r}_{j,z} \right) \left( \vec{l}_{j,z}^T \cdot \vec{p}_z(0) \right)  e^{ \lambda_j(z) t} \right]_{z=s}\, .
\end{equation}
The summation over $j$ complicates the time dependence of the generalized factorial cumulants.
In the long-time limit, however, the above expression becomes considerably simpler since the exponential factors suppress all terms of the sum except the ones with the largest real part $\tx{Re} \left[ \lambda_j (z) \right]$ for $z=s$.
For $z=1$ and systems with a unique stationary state, the dominant eigenvalue is $0$, i.e., all other eigenvalues have a negative real part.
Around $z=1$, the dominant eigenvalue, denoted by $\lambda_{\tx{max}} (z)$, remains real and the limit 
\begin{equation}
	\label{eq:longcum}
	c_{s,k}:= \lim_{t \to \infty}\frac{C_{s,k}(t)}{t} = \frac{\partial^k \lambda_{\tx{max}}(z)}{\partial z^k} \bigg| _{z=s},
\end{equation}
provides well-defined constants, referred to as scaled long-time (generalized factorial) cumulants.

These scaled long-time cumulants $c_{s,k}$ define the input information for the inverse counting statistics. Of course, in an experiment, the time $t$ is always finite and, moreover, $\lambda_{\tx{max}} (z)$ is not directly accessible. Therefore, one has to use scaled finite-time cumulants $C_{s,k}(t)/t$, with $C_{s,k}(t)$ obtained via equations~(\ref{generating_function}) and~(\ref{eq:Csk}) from the measured $P_N(t)$. For large $t$, the scaled cumulants become time independent such that $C_{s,k}(t)/t\approx c_{s,k}$.

If $s$ is chosen very negative, it may happen that the dominant eigenvalues are given by a complex-conjugated pair.
In this case, the limit is not well defined and the inverse counting-statistics procedure derived below cannot be applied.

\section{Inverse Counting Statistics}\label{sec:ics}
In the previous section, we have shown how to calculate for a given stochastic model defined by the matrix ${\bf W}_z$ (referred to as the generator of the stochastic system) the long-time (generalized factorial) cumulants.
Inverse counting statistics deals with the opposite problem: how much can we learn about the stochastic system if only a few numbers, namely the experimentally determined values of the scaled long-time cumulants (up to some order), are given?
To be more specific, we aim at the following properties of the stochastic system.
First, a very important feature of the stochastic system is the dimension $M$ of ${\bf W}_z$, i.e., the number of participating states in the stochastic process.
Furthermore, the coupling to the detector is described by powers of $z$ attached to some matrix elements of the generator. 
As a consequence, the characteristic polynomial $\tx{det}\left( \lambda \vec{1}-\vec{W}_{z}  \right)$ is of order~$m$ in $z$.
Thus, as a second feature, we identify the counting power~$m$.
We will show below that the values of the first $(m+1)M$ scaled long-time cumulants are enough to check compatibility with a stochastic system of dimension $M$ and the counting power~$m$ characterizing the coupling to the detector.

But, with inverse counting statistics, we can get much more.
From the $(m+1)M$ input parameters $c_{s,k}$ it is possible to determine the full spectrum of ${\bf W}_z$ , i.e., the full $z$-dependence of the eigenvalues $\lambda_j(z)$.
To appreciate how remarkable this statement is, let us remind that the input parameters are only a finite [$(m+1)M$] number of derivatives of only one eigenvalue $\lambda_{\tx{max}}$ at only one value of $z$, namely the arbitrarily chosen $s$.
From this rather restricted amount of information, we aim at reconstructing also the other eigenvalues different from $\lambda_{\tx{max}}$ at all values of $z$ different from $s$.
How is this possible and how does it work in practice?

To answer this question, we observe that the characteristic function of the generator~$\vec{W}_z$, 
\begin{equation}
\label{eq:chi}
	\chi(z,\lambda) = \tx{det}\left( \lambda \vec{1}-\vec{W}_{z}  \right)=\prod_{j=1}^M \left( \lambda - \lambda_j(z)  \right)\, ,
\end{equation}
is a polynomial both in $\lambda$ (of order $M$) and in $z$ (of order $m$).
The eigenvalues $\lambda_j(z)$, i.e., the zeros of the characteristic function, $\chi(z,\lambda_j(z))=0$, are, in general, nonanalytic functions in $z$.
The characteristic function $\chi(z,\lambda)$ itself, however, is a polynomial in $z$ and can, therefore, be written in the form
\begin{equation}
\label{eq:characteristic}
	\chi(z,\lambda) = \lambda^M + \sum_{\mu=0}^m \sum_{\nu=0}^{M-1}a_{\mu \nu}\, (z-s)^\mu \, \lambda^\nu \, ,
\end{equation}
where $s$ is the arbitrarily chosen parameter of the generalized factorial cumulants.
As a consequence, the ($s$-independent) characteristic function is fully determined by the $(m+1)M$ real (and $s$-dependent) coefficients $a_{\mu \nu}$.
This fixes all $z$-dependent (but $s$-independent) eigenvalues $\lambda_j(z)$ of the generator $\vec{W}_z$.
For this reason, $(m+1)M$ input parameters are enough to fully determine the spectrum of $\vec{W}_z$.

Suppose that $M$ and $m$ are already known (we will discuss below how this is done with the help of inverse counting statistics).
How do we get the spectrum of $\vec{W}_z$?
As input parameters we use the scaled generalized factorial cumulants $c_{s,k}$ for $k=0, \ldots, (m+1)M-1$ in the long-time limit. 
To determine the coefficients $a_{\mu\nu}$, we perform $l=0,\ldots, (m+1)M-1$ times a derivative of $\chi(z,\lambda (z)) \equiv 0$ with respect to $z$ and set $z=s$ afterwards.
For technical reasons, it is convenient to divide the resulting equation by $l!$.
Then, we arrive at the set of linear equations
\begin{equation}
\label{eq:linear}
	0 = A_{l,0M} + \sum_{\mu=0}^m \sum_{\nu=0}^{M-1}A_{l,\mu \nu} a_{\mu \nu} \, ,
\end{equation}
for $l=0, \ldots, (m+1)M-1$.
The coefficients $A_{l,\mu \nu}$, defined for nonnegative $l$, $\mu$, and $\nu$, are given by
\begin{equation}
A_{l,\mu \nu} = \left\{
\begin{array}{l c l}
	0 & \qquad \text{for} \, &l<\mu \\
	1 & \qquad \text{for} \, & l=\mu, \nu=0 \\
	0 & \qquad \text{for} \, & l>\mu, \nu=0
\end{array}
\right.
\end{equation}
and otherwise (i.e., $l\ge \mu$ together with $\nu\ge 1$) by
\begin{equation}\label{eq:As}
	A_{l,\mu \nu} = \sum_{\alpha_1 +\ldots + \alpha_\nu = l-\mu}
	 \frac{c_{s,\alpha_1}}{\alpha_1!}\cdot \frac{c_{s,\alpha_2}}{\alpha_2!}\cdots \frac{c_{s,\alpha_\nu}}{\alpha_\nu!}
\, .
\end{equation}
Obviously, $A_{l,\mu \nu}$ depends on $l$ and $\mu$ only via the difference $l-\mu$ (this was the reason of dividing by $l!$).
The multiple sum over the $\alpha$'s is constrained by $\alpha_1 +\ldots + \alpha_\nu = l-\mu$. An alternative expression for the $A_{l,\mu \nu}$ can be found in~\ref{sec:Apprandtel}.

To be explicit, let us write down all the terms that are relevant for the case $M=3$ and $m=2$.
We need $\nu$ up to $2$ and $l$ up to $8$.
For $\nu=0$ we get $A_{l,\mu 0} = \delta_{l\mu}$, for $\nu=1$ we have $A_{l,\mu 1} = c_{s,l-\mu}/(l-\mu)!$, and for $\nu=2$ we find
\begin{eqnarray}
	A_{\mu,\mu 2} &=& c_{s,0}^2 \, ,
\nonumber \\
	A_{\mu+1,\mu 2} &=& 2 c_{s,0}c_{s,1} \, ,
\nonumber \\
	A_{\mu+2,\mu 2} &=& c_{s,0}c_{s,2} + c_{s,1}^2 \, ,
\nonumber \\
	A_{\mu+3,\mu 2} &=& \frac{c_{s,0}c_{s,3}}{3} + c_{s,1}c_{s,2} \, ,
\nonumber \\
	A_{\mu+4,\mu 2} &=& \frac{c_{s,0}c_{s,4}}{12} + \frac{c_{s,1}c_{s,3}}{3} + \frac{c_{s,2}c_{s,2}}{4} \, ,
\nonumber \\
	A_{\mu+5,\mu 2} &=& \frac{c_{s,0}c_{s,5}}{60} + \frac{c_{s,1}c_{s,4}}{12} + \frac{c_{s,2}c_{s,3}}{6} \, ,
\nonumber \\
	A_{\mu+6,\mu 2} &=& \frac{c_{s,0}c_{s,6}}{360} + \frac{c_{s,1}c_{s,5}}{60} + \frac{c_{s,2}c_{s,4}}{24} + \frac{c_{s,3}c_{s,3}}{36} \, ,
\nonumber \\
	A_{\mu+7,\mu 2} &=& \frac{c_{s,0}c_{s,7}}{2520} + \frac{c_{s,1}c_{s,6}}{360} + \frac{c_{s,2}c_{s,5}}{120} + \frac{c_{s,3}c_{s,4}}{72} \, ,
\nonumber \\
	A_{\mu+8,\mu 2} &=& \frac{c_{s,0}c_{s,8}}{20160} + \frac{c_{s,1}c_{s,7}}{2520} + \frac{c_{s,2}c_{s,6}}{720} + \frac{c_{s,3}c_{s,5}}{360} + \frac{c_{s,4}c_{s,4}}{576} \, . 
\nonumber 
\end{eqnarray}

To obtain the full spectrum $\{ \lambda_j(z) \}$, one first solves the set of linear equations~(\ref{eq:linear}) for $a_{\mu\nu}$.
Second, the result for $a_{\mu\nu}$ is inserted into equation~(\ref{eq:characteristic}) to get the characteristic function.
Finally, the zeros of the characteristic function are determined.
It all works because the characteristic function is fully determined by a finite number of coefficients~$a_{\mu\nu}$ only.

It is important to remark that, in general, there is no guarantee that the set of linear equations~(\ref{eq:linear}) provides a unique solution.
There may be special situations (we will discuss such a case below) in which the generator $\vec{W}_z$ is separable in the sense that its characteristic function can be written as a product of two polynomials, the first one of order $m'$ and $M'$ in $z$ and $\lambda$, the second one of order $m-m'$ and $M-M'$.
Of course, it is trivial that the characteristic function can always be written as a product of two polynomials in $\lambda$, but separability requires that, in addition, the two polynomials in $\lambda$ are polynomials in $z$ as well. 
For separable generators, inverse counting statistics has only access to the part of the spectrum to which the eigenvalue $\lambda_{\tx{max}}(z)$ with the largest real part belongs, i.e., the effective problem has reduced dimensions $M'$ and $m'$.

The remaining question to be answered is how to determine the dimensions $M$ and $m$ of the stochastic system.
In the spirit of~\onlinecite{bruderer_inverse_2014}, one may suggest that after having determined $\lambda_{\tx{max}}(z)$ from the first $(m+1)M$ scaled long-time cumulants $c_{s,k}$ one can calculate the $c_{s,k}$ of higher order and compare them with experimentally measured values~$C_{s,k}(t)/t$ in the long-time limit.
If $M$ and $m$ were chosen correctly, then one expects a coincidence of calculated and measured values. 
This has, however, two downsides.
First, measuring cumulants of increasingly higher order may become more and more difficult. 
Second, comparing just numbers to establish a consistency criterion may be of only limited significance in view of experimentally unavoidable noise.

With the use of generalized factorial cumulant, however, we can do much better.
Remember that the parameter $s$ was chosen arbitrarily and the result, i.e., the spectrum of the generator, must be independent of this parameter~$s$.
Therefore, we can use the very same measured time trace of the detector to determine scaled cumulants~$C_{s,k}(t)/t$ in the long-time limit for different values of $s$ and, afterwards, run the inverse counting-statistics procedure for each $s$. 
If $M$ and $m$ were chosen correctly, then the {\it full} $z$-dependence of the {\it full} spectrum should be independent of the choice of $s$.
This establishes a much stronger consistency check than comparing just a few numbers.

In the following sections~\ref{sec:QDZee} and \ref{sec:SEBAndreev}, we illustrate the inverse counting-statistics procedure for two model systems: a single-level quantum dot in a Zeeman field and a single-electron box subjected to sequential and Andreev tunneling.

\section{Single-level quantum dot in a Zeeman field}\label{sec:QDZee}
\begin{figure}[t]
\begin{center}
\includegraphics[scale=1.10]{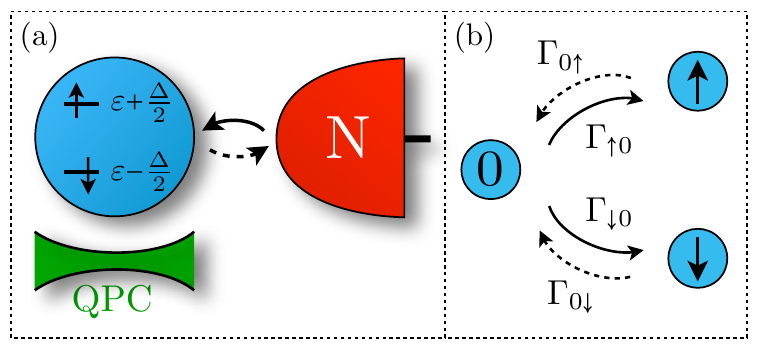}
\caption{
	(Color online) (a) A single-level quantum dot subjected to a Zeeman field is tunnel coupled to one normal lead. (b) Sketch of the states and transition rates. Dashed arrows indicate the counted transitions.
		}
\label{fig2}
\end{center}
\end{figure}
The first model system is depicted in figure~\ref{fig2}(a). A single-level quantum dot is weakly tunnel coupled to one normal-state metallic lead N and subjected to a magnetic field. The orbital level $\varepsilon$ measured relative to lead's electrochemical potential is splitted by the Zeeman energy $\Delta$ into $\varepsilon_\sigma=\varepsilon \pm\Delta/2$. The positive (negative) sign applies to a spin $\sigma=\uparrow (\downarrow)$ electron on the quantum dot. The current through a nearby quantum point contact is sensitive to the dot charge which allows to monitor the charge transfer between dot and lead as function of time. The empty dot $0$ can be occupied by a spin $\sg$ electron with the sequential tunneling rate $\Gamma_{\sg 0}=\Gamma f(\varepsilon_\sg)$ given by Fermi's golden rule. The reverse transition occurs with the rate $\Gamma_{0 \sg }=\Gamma \left[ 1- f(\varepsilon_\sg) \right]$, where $f(\varepsilon_\sg) = \left[ 1+\exp (\varepsilon_\sg / k_\text{B}T) \right]^{-1}$ is the Fermi function. The temperature $T$ is as large that both $\Gamma_{\sg 0}$ and $\Gamma_{0 \sg}$ are nonvanishing, but transitions to higher charge states are negligible due to charging energy.

The stochastic system is depicted in figure~\ref{fig2}(b). Its generator is given by
\begin{equation}\label{eq:Wzexampledot}
	\vec{W}_{z} =\begin{pmatrix} -\Gm_{\uparrow 0}- \Gm_{\downarrow 0}& z \Gm_{0\uparrow} & z \Gm_{0\downarrow}\\ \Gm_{\uparrow0} & -\Gm_{0\uparrow}&  0 \\ 	\Gm_{\downarrow0} & 0 & -\Gm_{0\downarrow} \end{pmatrix} \, .
\end{equation}
Each counting factor $z$ in the upper-right off-diagonal matrix elements correspond to counting an electron leaving the dot.
The characteristic function is a polynomial of order $M=3$ in $\lambda$ and of order $m=1$ in~$z$.

The case of vanishing magnetic field $\Delta=0$, however, is special because of spin degeneracy $\varepsilon_\uparrow=\varepsilon_\downarrow$, and only two different rates $\Gm_{\uparrow 0}=\Gm_{\downarrow 0}=\Gm_{10}$ and $\Gm_{0 \uparrow}=\Gm_{0 \downarrow}=\Gm_{01}$ appear.
As a consequence, the characteristic function becomes separable,
\begin{eqnarray}
	\chi (z,\lambda) &=& \chi_{1,2} (z,\lambda) \cdot \chi_3 (\lambda)\, ,
\\
	\chi_{1,2} (z,\lambda) &=& \lambda^2 + (\Gm_{01}+2\Gm_{10})\lambda - 2 (1-z) \Gm_{01}\Gm_{10}\, ,
\\
	\chi_3(\lambda) &=& \lambda + \Gm_{01} \, .
\end{eqnarray}
The first factor is a polynomial of order $M=2$ in $\lambda$ and of order $m=1$ in $z$, while the second factor is of order $M=1$ and independent of $z$. Due to the $z$-independency, the second factor does not influence counting statistics and thus can not be detected anymore. The electron transfer can be completely described by a spinless orbital which is occupied with rate $2\Gm_{10}$ and emptied with rate $\Gm_{01}$. The factor 2 reflects the spin degeneracy~\cite{beckel_asymmetry_2014}. Note that a separable characteristic function does not always separate in a $z$-dependent and $z$-independent factor, an example will be given in section~\ref{sec:SEBAndreev}.

\subsection{Nonvanishing magnetic field}
We start with discussing the case of nonvanishing magnetic field for which we choose $\varepsilon=- k_\tx{B}T$ and $\Delta=k_\tx{B}T/2$. As unit of time we chose the inverse of $\Gm = \Gm_{\sg 0} + \Gamma_{0 \sg }$.
The input information for the inverse counting statistics (for $M=3$ and $m=1$) is given by the scaled long-time generalized factorial cumulants $c_{s,k}$ from order $0$ up to order $(m+1)M-1=5$.
Since in experiments, the measurement time is always finite, we do not take as input parameters the exact scaled long-time cumulants of the defined model but calculate, instead, the scaled finite-time cumulants $C_{s,k}(t)/t$ at some large but finite time~$t$. Hence, the scaled cumulants are close but not identical to the exact values in the long-time limit. For the exact long-time scaled cumulants, a value for the dimension~$M$ that was assumed to be too large, can be immediately identified by trying to solve the system of linear equations~(\ref{eq:linear}). Then, no unique solution for every $a_{\mu \nu}$ is obtained, i.e., the linear equations are not independent from each other. However, in the following, we stick to the experimental situation that, due to the finite measuring time or due to experimental noise, a unique solution for the system of linear equations~(\ref{eq:linear}) is even obtained if $M$ is assumed too large.

\begin{figure}[t]
\includegraphics[scale=1.00]{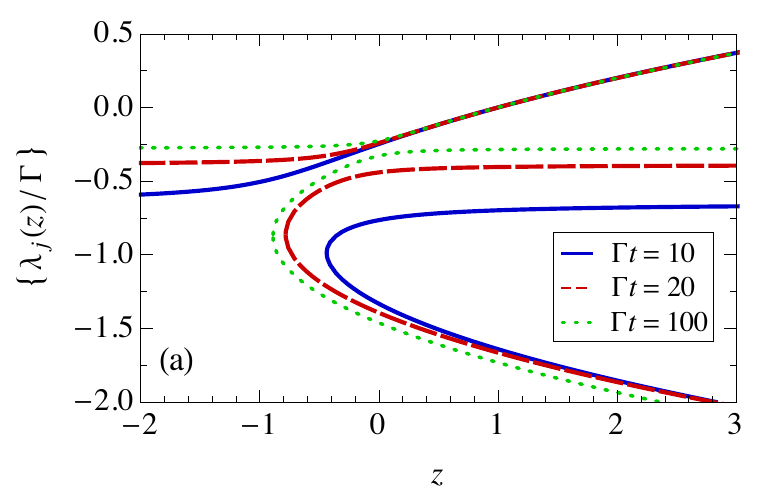}
\includegraphics[scale=1.00]{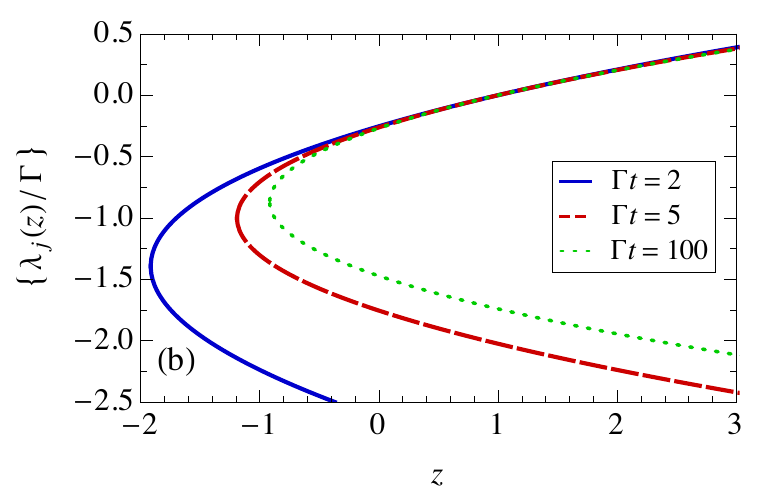}
\caption{
	(Color online) Eigenvalue spectrum obtained via inverse counting statistics for a single-level quantum dot $\varepsilon=-k_\tx{B}T$ (a) inside a magnetic field $\Delta=k_\tx{B}T/2$ and (b) without magnetic field. The input data are scaled cumulants $C_{s,k}(t)/t$ for finite times $t$. Only in the long-time limit $\Gm t \gtrsim 100$, the output $\{ \lambda_j \}$ is indeed the eigenvalue spectrum of the generator~(\ref{eq:Wzexampledot}). Values $\lambda_j $ with a finite imaginary part (occurring for very negative $z$) are not depicted.
		}
\label{fig3}
\end{figure}

First, we determine the eigenvalue spectrum from the inverse counting statistics performed at $s=1$ and assuming the correct values $M=3$ and $m=1$.
As input parameter, we take the calculated scaled generalized factorial cumulants $C_{s,k}(t)/t$ at times $\Gm t=10, 20, 100$.
The result is shown in figure~\ref{fig3}(a). For $\Gm t=100$, the long-time limit is reached and no difference to the exact eigenvalues can be recognized anymore.

Next, we demonstrate the consistency check for the dimensions $M$ and $m$.
For this, we check the required $s$-independence of the eigenvalue spectrum.
In the following, we always use as input information the calculated scaled cumulants at $\Gm t= 6000$ to ensure convergence to the asymptotic long-time behavior not only for $s=1$ but also for other~$s$.  

\begin{figure}[t]
\begin{center}
{\includegraphics[scale=1.00]{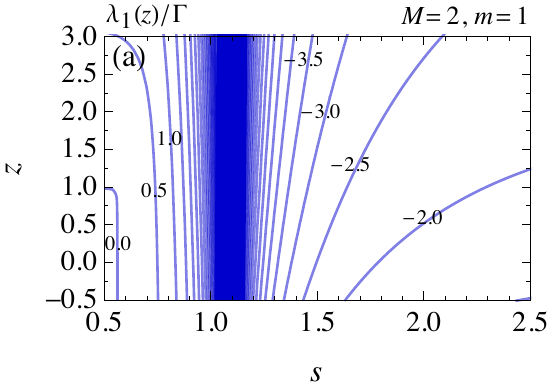}}
{\includegraphics[scale=1.00]{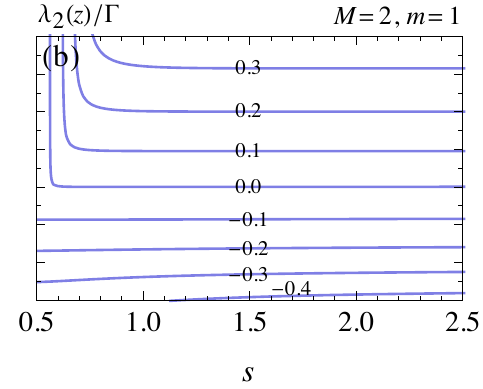}}
{\includegraphics[scale=1.00]{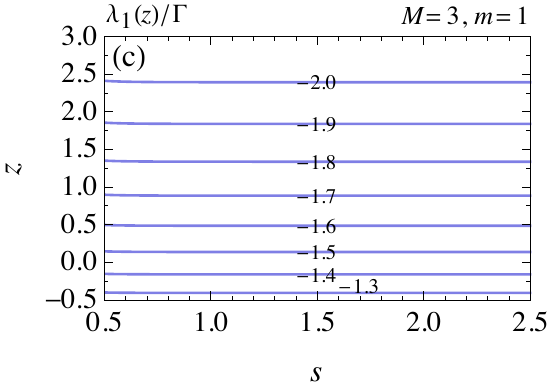}}
{\includegraphics[scale=1.00]{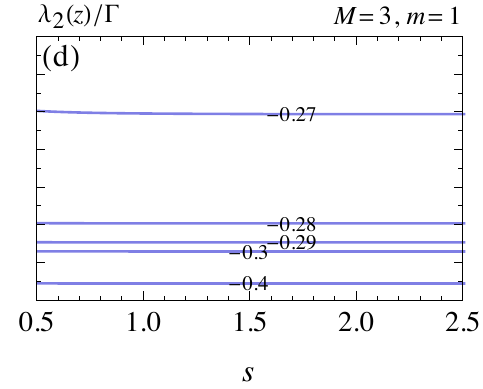}}
{\includegraphics[scale=1.00]{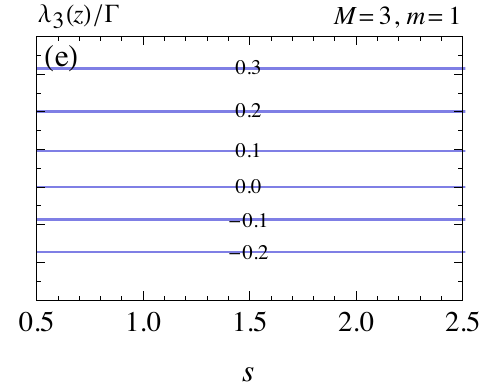}}
\caption{
	Consistency check of the number of system states~$M$ and the counting power~$m$ for the single-level quantum dot $\varepsilon=- k_\tx{B}T$ in a magnetic field $\Delta=k_\tx{B}T/2$ for time $\Gm t=6000$. Assuming different values for $M$ and $m$, contour lines of the resulting $j=1,\dots,M$ eigenvalues $\lambda_j(z) / \, \Gm$ are depicted. The eigenvalues for $M=2$, $m=1$ in (a), (b) show a strong $s$-dependence, the ones for $M=3$, $m=1$ in (c), (d), (e) are $s$-independent.
	}
\label{fig4}
\end{center}
\end{figure}

To show simultaneously both the $z$- and the $s$-dependence of the eigenvalues, we plot in the following figures the contour lines for different selected values of $\lambda$ (in units of $\Gm$). 
Horizontal contour lines indicate that the eigenvalues are independent of $s$. If {\it all} eigenvalues are $s$-independent, then, the assumed dimensions $M$ and $m$ are compatible with the input data.
In figure~\ref{fig4}(a) and (b), we show the result for the choice $M=2$ and $m=1$.
Since the dimension $M$ of the stochastic systems is taken too small, the resulting eigenvalues display a strong $s$-dependence, especially the eigenvalue~$\lambda_1$.
However, if we take the proper values $M=3$ and $m=1$, see figure~\ref{fig4}(c), (d), (e), we get $s$-independent results. The order $M=3$ and $m=1$ are lower bounds for the system's dimensions.

\subsection{Vanishing magnetic field}\label{sec:outCBlong}
\begin{figure*}[t]
\begin{center}
{\includegraphics[scale=1.00]{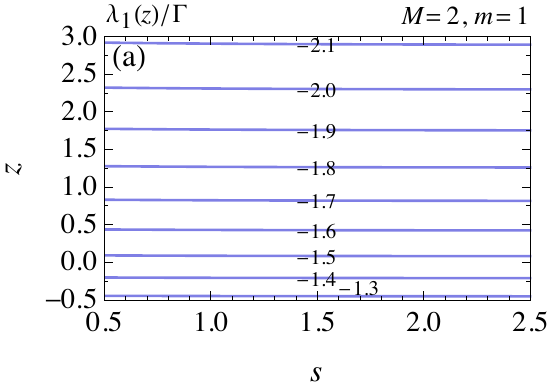}}
{\includegraphics[scale=1.00]{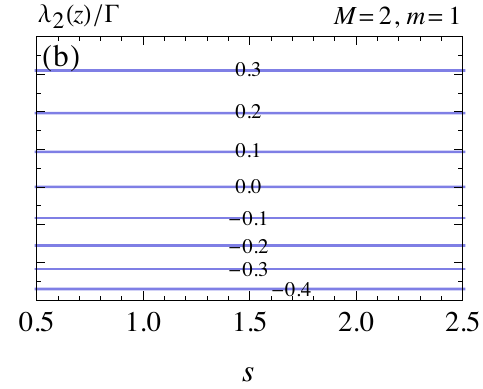}}
\caption{
	Consistency check of the number of system states~$M$ and the counting power~$m$ for the single-level quantum dot $\varepsilon=- k_\tx{B}T$ without magnetic field for time $\Gm t=100$. Assuming $M=2$ and $m=1$, contour lines of the resulting two eigenvalues $\lambda_1(z) / \, \Gm$ and $ \lambda_2(z) / \, \Gm$ are $s$-independent.
	}
\label{fig5}
\end{center}
\end{figure*}

We now turn to the case of vanishing magnetic field $\Delta=0$, for which the characteristic function is separable, i.e., the characteristic function is a product of two polynomials, one of order $M=2$ and $m=1$ in $\lambda$ and $z$, and the other one is of order $M=1$ and independent of $z$. The latter polynomial, i.e., the third dimension, does not influence transport anymore because the system can be described in a spineless model with $M=2$ and $m=1$.

Performing the inverse counting statistics at $s=1$ with the correct values $M=2$ and $m=1$ gives the spectrum depicted in figure~\ref{fig3}(b). The input information is given by the scaled finite-time cumulants $C_{s,k}(t)/t$ from order $0$ to $3$. For $\Gm t=100$, the long-time limit is reached and no difference to the exact eigenvalues can be recognized anymore.

For the consistency check of the dimensions $M$ and $m$, we use as input information the calculated scaled cumulants at $\Gm t= 100$. Horizontal contour lines in figure~\ref{fig5}(a), (b) indicate that the eigenvalues are independent of $s$, i.e., the assumed dimensions $M=2$ and $m=1$ are compatible with the input data. In contrast for the choice $M=3$ and $m=1$, we obtain no $s$-independent spectrum of eigenvalues. The dimension $M=3$ is too large.

For completeness, we discuss in~\ref{sec:deltaapprox0} how close the magnetic field has to be tuned to $\Delta=0$ in order to observe the discussed behavior.

\section{Sequential and Andreev tunneling in a single-electron box}\label{sec:SEBAndreev}
\begin{figure}[t]
\begin{center}
\includegraphics[scale=1.10]{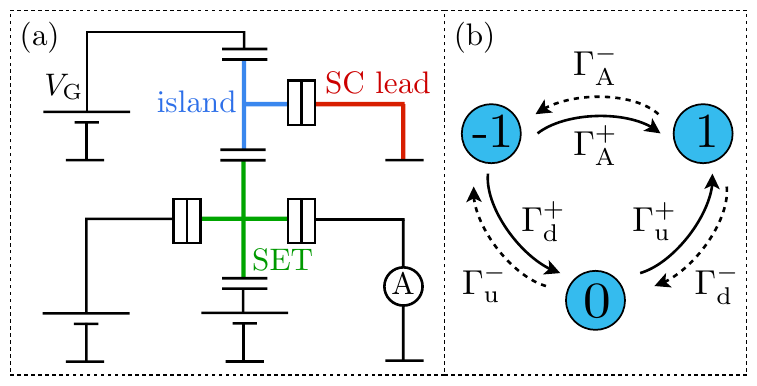}
\caption{
	(Color online) (a) A normal-state metallic island (blue) weakly tunnel coupled to one superconducting lead (red). Via the gate voltage $V_\tx{G}$, the gate charge $n_\tx{G}$ of the island can be tuned. By means of a current measurement trough a single-electron transistor (green), the number ob excess electrons $n$ on the island is obtained as function of time. (b) Sketch of the states and transition rates. Dashed arrows indicate the counted transitions.
		}
\label{fig6}
\end{center}
\end{figure}
The second example illustrating the inverse counting-statistics procedure is a model system that has been already experimentally realized in~\onlinecite{saira_environmentally_2010, maisi_real_2011, maisi_full_2014}. The set up is depicted in figure~\ref{fig6}(a). A single-electron box (SEB) is formed by one superconducting lead weakly coupled (characterized by the tunnel resistance $R_\tx{T}$) to a normal-state metallic island. The energy required to bring $n$ excess electrons on the island is $E_C(n-n_\tx{G})^2$. By applying a voltage $V_\tx{G}$ to a gate electrode, the gate charge is tuned near $n_\tx{G}=0$. The charge $n$ on the island is monitored by an electrostatically coupled single-electron transistor (SET): each value of $n$ results in a characteristic value of the current through the SET. Due to a finite temperature, transitions $n=0 \to \pm 1$ with the rate $\Gm_{\tx{u}}^\pm$ are possible (but temperature is small enough so that transitions to further charge states $-2$ and $2$ are negligible). If the island is in one of these excited states, Andreev tunneling $n=\pm1 \to \mp1$ with the rate $\Gm_{\tx{A}}^\mp$ is possible until the island relaxes back to the ground state $n=\pm1 \to 0$ with rate $\Gm_{\tx{d}}^\mp$. 

The stochastic system is depicted in figure~\ref{fig6}(b).
Its generator is given by
\begin{equation}\label{eq:Wzexample}
	\vec{W}_{z} =\begin{pmatrix} -\Gm_{\tx A}^+ - \Gm_{\tx d}^+ & z \Gm_{\tx u}^- & z^2 \Gm_{\tx A}^-\\ \Gm_{\tx d}^+& -\Gm_{\tx u}^+ - \Gm_{\tx u}^-& z \Gm_{\tx d}^-\\ 	\Gm_{\tx A}^+&\Gm_{\tx u}^+ & -\Gm_{\tx A}^- -\Gm_{\tx d}^- \end{pmatrix} \, .
\end{equation}
Each counting factor $z$ in the upper-right off-diagonal matrix elements correspond to counting an electron leaving the island. Since an Andreev-tunneling process transfers two electrons, $\Gm_{\tx A}^-$ is multiplied with the square $z^2$ instead of a single~$z$. This counting procedure is, especially, reasonable if the detector can not resolve two separate transitions $-1 \to 0$ followed directly by $0 \to +1$ from the Andreev-tunneling process $-1 \to +1$. For the chosen counting procedure, theses transitions need not to be distinguished.

If transitions are counted in which the number of electrons on the island is increased, the counting factors are removed from the upper-right off-diagonal matrix elements and put, instead, to the lower-left terms.
The cumulants are identical for both counting procedures in the long-time limit. However, for finite times and asymmetric transition rates, the cumulants can differ, which indicates a violation of detailed balance~\cite{stegmann_violation_2016}.

The characteristic function is a polynomial of order $M=3$ in $\lambda$ and of order $m=2$ in~$z$.
In general, the inverse counting statistics should deliver the $z$-dependence of all three eigenvalues.

The symmetry point $n_\tx{G}=0$, however, is special.
At this point, there are only three different rates $\Gm_{\tx A} =\Gm_{\tx A}^+ = \Gm_{\tx A}^-$, $\Gm_{\tx d} =\Gm_{\tx d}^+ = \Gm_{\tx d}^-$, and $\Gm_{\tx u} =\Gm_{\tx u}^+ = \Gm_{\tx u}^-$.
As a consequence, the characteristic function becomes separable,
\begin{eqnarray}
	\chi (z,\lambda) &=& \chi_{1,2} (z,\lambda) \cdot \chi_3 (z,\lambda)\, ,
\\
	\chi_{1,2} (z,\lambda) &=& \lambda^2 + [ (1-z) \Gm_{\tx A} + \Gm_{\tx d} + 2\Gm_{\tx u}] \lambda + 2(1-z)\Gm_{\tx u} (\Gm_{\tx A} + \Gm_{\tx d})\, ,
\\
	\chi_3 (z,\lambda) &=& \lambda + (1+z) \Gm_{\tx A} + \Gm_{\tx d} \, .
\end{eqnarray}
The first factor is a polynomial of order $M=2$ in $\lambda$ and of order $m=1$ in $z$, while the second one is of order $M=1$ and $m=1$.
Depending on the choice of $s$, the inverse counting statistics will only provide the two eigenvalues of the first or the single eigenvalue of the second factor.
Away from the symmetry point, $n_\tx{G} \neq 0$, the generator is not separable.
\begin{figure}[t]
\includegraphics[scale=1.00]{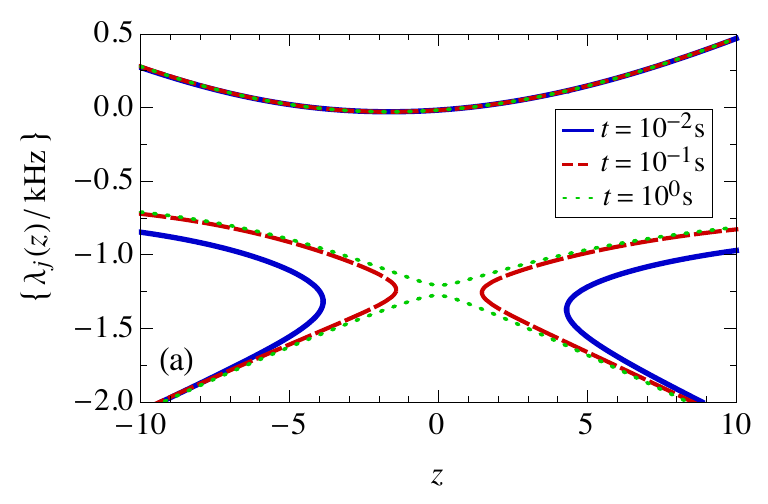}
\includegraphics[scale=1.00]{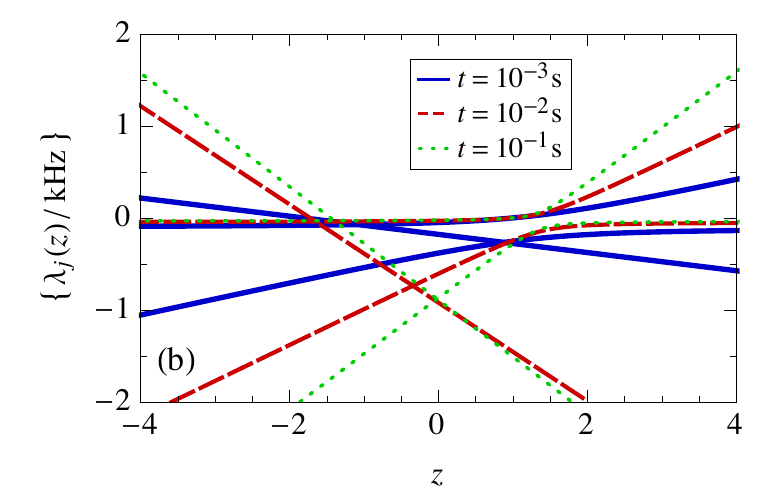}
\caption{
	(Color online) Eigenvalue spectrum obtained via inverse counting statistics for a normal-state metallic island weakly tunnel coupled to one superconducting lead. The gate charge is (a) $n_\tx{G}=0.09$ and (b) $n_\tx{G}=0.00$. The input data are scaled cumulants $C_{s,k}(t)/t$ for finite times $t$. Only in the long-time limit, (a) $t\gtrsim10^0 \, \tx{s}$ and (b) $t\gtrsim10^{-1} \, \tx{s}$, the output $\{ \lambda_j \}$ is indeed the eigenvalue spectrum of the generator~(\ref{eq:Wzexample}). Values $\lambda_j $ with a finite imaginary part (occurring for very negative $z$) are not depicted.
		}
\label{fig7}
\end{figure}

Instead of calculating the tunneling rates $\Gm_\tx{u}^\pm$, $\Gm_\tx{d}^\pm$, and $\Gm_\tx{A}^\pm$ in the presence of an electromagnetic environment~\cite{saira_environmentally_2010, pekola_environment_2010}, we rely on experimentally measured rates for $n_\tx{G}=0.09$ in~\onlinecite{maisi_real_2011} and $n_\tx{G}=0.00$ in~\onlinecite{maisi_full_2014}. In the former case, the experimental parameters are $n_\tx{G}=0.09$, $E_\tx{C}=43 \mu \tx{eV}$, $\Delta=216 \mu \tx{eV}$, and $R_\tx{T}=2000\  \tx{k} \Omega$ at $60\, \tx{mK}$ temperature. The measured rates are $\Gm_\tx{u}^+=10.5\ \tx{Hz}$, $\Gm_\tx{u}^-=7.2\ \tx{Hz}$, $\Gm_\tx{d}^+=1270\ \tx{Hz}$, $\Gm_\tx{d}^-=730\ \tx{Hz}$, $\Gm_\tx{A}^+=460\ \tx{Hz}$, and $\Gm_\tx{A}^-=23.0\ \tx{Hz}$. In the latter case, the experimental parameters are $n_\tx{G}=0.00$, $E_\tx{C}=40 \mu \tx{eV}$, $\Delta=210 \mu \tx{eV}$, and $R_\tx{T}=490\  \tx{k} \Omega$ at $50\, \tx{mK}$ temperature. The rates are $\Gm_\tx{u}^+=\Gm_\tx{u}^-=\Gm_\tx{u}=12\ \tx{Hz}$, $\Gm_\tx{d}^+=\Gm_\tx{d}^-=\Gm_\tx{d}=252\ \tx{Hz}$, and $\Gm_\tx{A}^+=\Gm_\tx{A}^-=\Gm_\tx{A}=615\ \tx{Hz}$.

\subsection{Non-symmetric case}
We start with discussing the generic, non-symmetric case, for which we choose the $n_\tx{G}=0.09$.
The input information for the inverse counting statistics (for $M=3$ and $m=2$) is given by the scaled long-time generalized factorial cumulants $c_{s,k}$ from order $0$ up to order $(m+1)M-1=8$.
Similar to the case of the single-level quantum dot discussed in section~\ref{sec:QDZee}, we take into account that the measurement time is always finite in an experiment and do not take as input parameters the exact scaled long-time cumulants of the defined model but calculate, instead, the scaled cumulants $C_{s,k}(t)/t$ at some large but finite time.

First, we determine the eigenvalue spectrum from the inverse counting statistics performed at $s=1$ and assuming the correct values $M=3$ and $m=2$.
As input parameter we take the calculated scaled generalized factorial cumulants at times $t=0.01, 0.1, 1\, \tx{s}$. The result is shown in figure~\ref{fig7}(a). For $t=1\, \tx{s}$ no difference to the exact eigenvalues can be recognized anymore.
\begin{figure*}[t]
\begin{center}
{\includegraphics[scale=1.00]{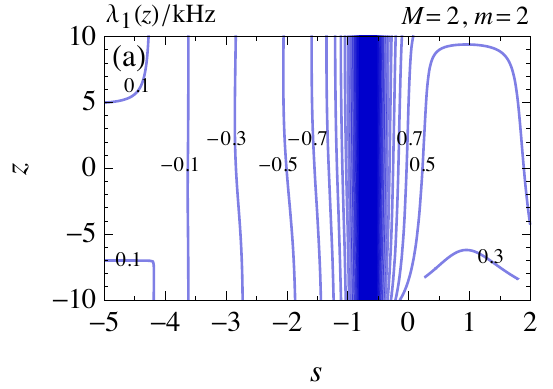}}
{\includegraphics[scale=1.00]{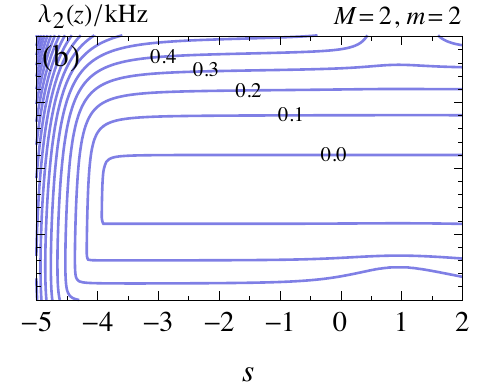}}\\
{\includegraphics[scale=1.00]{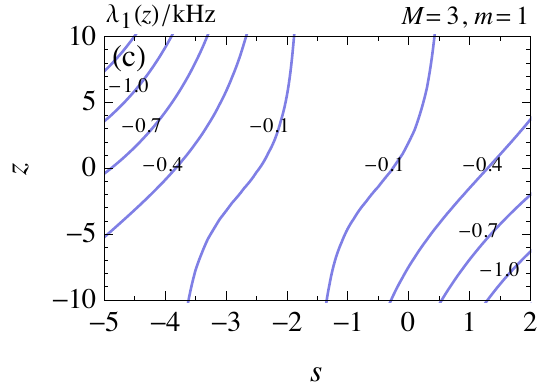}}
{\includegraphics[scale=1.00]{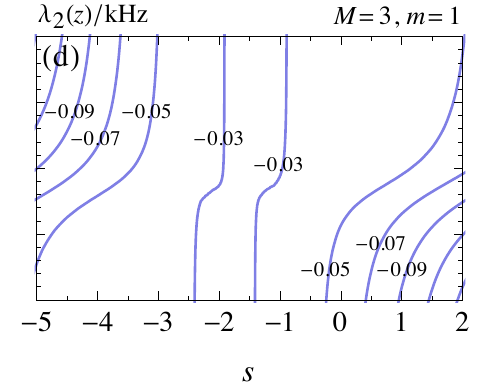}}
{\includegraphics[scale=1.00]{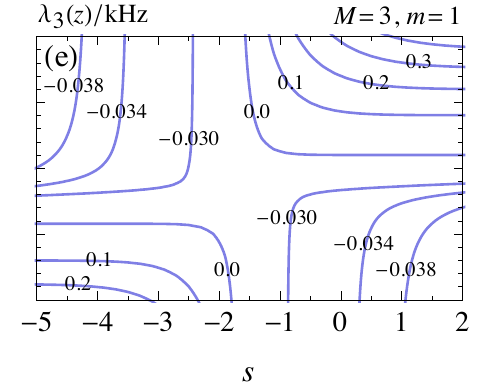}}\\
{\includegraphics[scale=1.00]{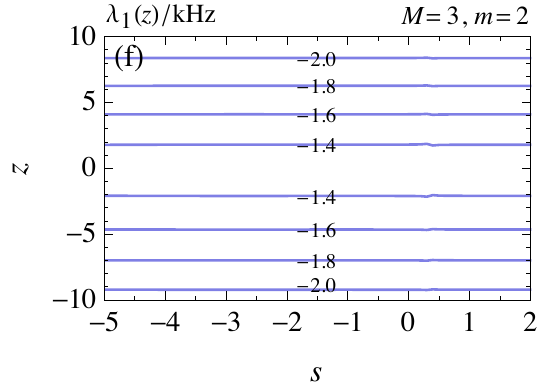}}
{\includegraphics[scale=1.00]{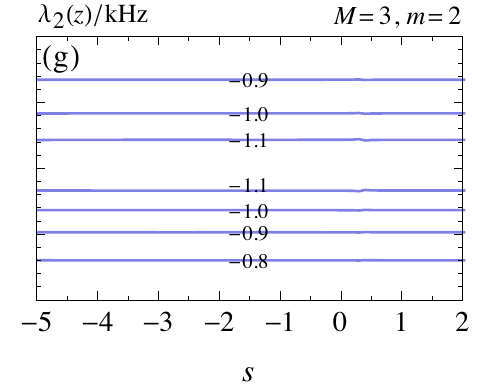}}
{\includegraphics[scale=1.00]{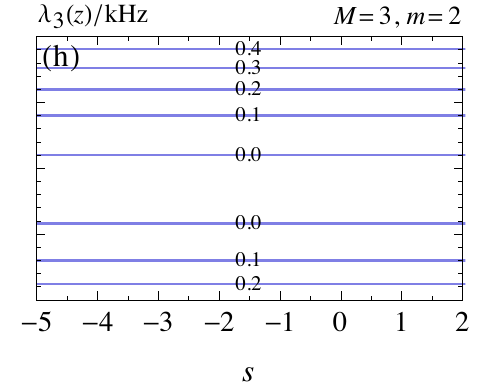}}

\caption{
		Consistency check of the number of system states~$M$ and the counting power~$m$ for the single-electron box with nonvanishing gate charge $n_\tx{G}=0.09$ for time $t=10\, \tx{s}$. Assuming different values for $M$ and $m$, contour lines of the resulting $j=1,\dots,M$ eigenvalues $ \lambda_j(z) / \, \tx{kHz}$ are depicted. The eigenvalues for $M=2$, $m=2$ in (a), (b) are strongly $s$-dependent, for $M=3$, $m=1$ in (c), (d), (e) as well, but the eigenvalues for $M=3$, $m=2$ in (f), (g), (h) are $s$-independent.
	}
\label{fig8}
\end{center}
\end{figure*}

Next, we demonstrate the consistency check for the dimensions $M$ and $m$. For the right values, all eigenvalues must be  $s$-independent. In the following, we always use as input information the calculated scaled cumulants at $t= 10\, \tx{s}$.
To show simultaneously both the $z$- and the $s$-dependence of the eigenvalues, we plot in the following figures contour lines for $\lambda$ (in units of kHz). 
Horizontal contour lines indicate that the eigenvalues are independent of $s$, i.e., the assumed dimensions $M$ and $m$ are compatible with the input data.

In figure~\ref{fig8}(a), (b), we show the result for the choice $M=2$ and $m=2$.
Since the dimension $M$ of the stochastic systems is taken too small, the resulting eigenvalues display a strong $s$-dependence.
The same holds true for the choice $M=3$ and $m=1$, see figure~\ref{fig8}(c), (d), (e).
In this case, the counting power $m$ characterizing the coupling to the detector is taken too small, and, again, the resulting eigenvalues are $s$-dependent.
Only if we take the proper values $M=3$ and $m=2$, see figure~\ref{fig8}(f), (g), (h), we get $s$-independent results.

\subsection{Symmetric case}\label{sec:inCBlong}
\begin{figure*}[t]
\begin{center}
{\includegraphics[scale=1.00]{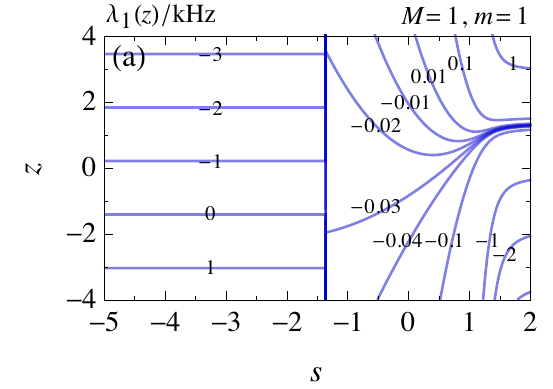}}
{\includegraphics[scale=1.00]{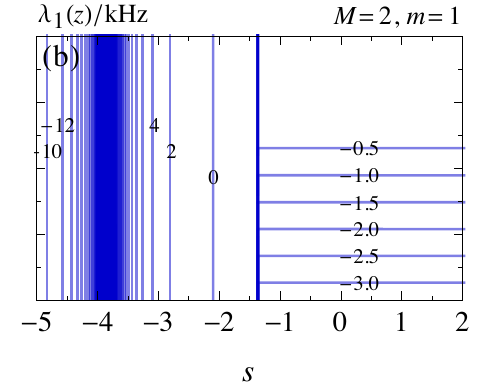}}
{\includegraphics[scale=1.00]{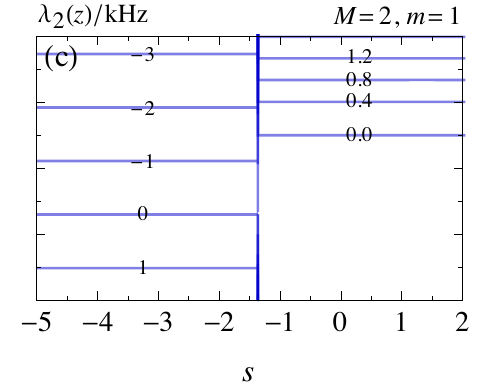}}
\caption{
	Consistency check of the number of system states~$M$ and the counting power~$m$ for the single-electron box with vanishing gate charge $n_\tx{G}=0$ for time $t=20\, \tx{s}$. Assuming different values for $M$ and $m$, contour lines of the resulting $j=1,\dots,M$ eigenvalues $ \lambda_j(z) / \, \tx{kHz}$ are depicted. The eigenvalue for $M=1$, $m=1$ in (a) and the two eigenvalues for $M=2$, $m=1$ in (b), (c).}
\label{fig9}
\end{center}
\end{figure*}

We now turn to the symmetric case, $n_\tx{G}=0$, for which the characteristic function is separable, i.e., the characteristic function is a product of two polynomials, one of order $M=2$ and $m=1$ in $\lambda$ and $z$, and the other one of order $M=1$ and $m=1$.
The choice of $s$ determines whether the eigenvalue with the largest real part is a zero of the first or the second factor and, therefore, which and how many of the eigenvalues are accessible via the inverse counting statistics.
If we choose $s=1$ (corresponding to factorial cumulants), then we obtain only two of the three eigenvalues. 
For small $s< -1.36$ (for figure~\ref{fig7}(b), we choose $s=-2$), we get only the third one. The number of required cumulants is 4 in the former and 2 in the latter case. The resulting spectrum of all three eigenvalues is depicted in figure~\ref{fig7}(b) for the times $t=0.001, 0.01, 0.1\, \tx{s}$. For $t=0.1\, \tx{s}$, no difference to the exact eigenvalues can be recognized anymore.

The consistency check of the dimensions $M$ and $m$ (for $t=20\, \tx{s}$) is depicted in figure~\ref{fig9}(a) for $M=1$ and $m=1$ and in figure~\ref{fig9}(b), (c) for $M=2$ and $m=1$.
If we perform the inverse counting statistics around $s=1$ (or for any $s>-1.36$), then we conclude that the dimension $M=1$ is too small (no horizontal contour lines in figure~\ref{fig9}(a) for $s>-1.36$), but $M=2$ seems to be sufficient (horizontal contour lines in Figs.~\ref{fig9}(b) and (c) for $s>-1.36$).
Thus, by employing only factorial cumulants ($s=1$) one may be tempted to conclude that the dimension of the stochastic system is $M=2$ only.
If, on the other hand, inverse counting statistics is also done for $s<-1.36$, then the horizontal contour lines in figure~\ref{fig9}(a) indicate that there is another eigenvalue.
Since the obtained $z$-dependent eigenvalues are different from each other, we conclude that there must be, in total, three eigenvalues.

For completeness, we estimate in~\ref{sec:ngapprox0} how close $n_\tx{G}$ has to be tuned to zero in order to observe the discussed behavior.

\section{Inverse-counting-statistics manual}\label{sec:summary}
For practical use, we summarize the inverse-counting-statistics procedure introduced in this paper in the following step-by-step manual:

\begin{itemize}
\item[(i)] Some positive integer values for the number of system states~$M$ and the counting power~$m$ are assumed.

\item[(ii)] From the measured probability distribution $P_N(t)$, the scaled {\it finite time} cumulants~$C_{s,k}(t)/t$ of order $k=0,\dots, (m+1)M-1$ for some real $s$ are calculated, see equations~$({\ref{generating_function}})$ and $(\ref{eq:Csk})$. The time $t$ is chosen large enough for $C_{s,k}(t)/t$ to become time independent such that $C_{s,k}(t)/t\approx c_{s,k}$.

\item[(iii)] From the~$c_{s,k}$ the set of linear equations~(\ref{eq:linear}) is derived.

\item[(iv)] Solving this set for the coefficients $a_{\mu \nu}$ yields the characteristic function $\chi(z,\lambda)$ defined in equation~(\ref{eq:chi}).

\item[(v)] Solving $\chi(z,\lambda_j)=0$ yields the zeros $\lambda_j(z)$.

\end{itemize}

If the $z$-dependence of {\it each} zero $\lambda_j(z)$ is independent of the parameter $s$ chosen in step (ii), the values assumed in step (i) are lower bounds for $M$ and $m$. Moreover, the $\lambda_j(z)$ are, indeed, eigenvalues of the system's generator $\vec{W}_z$ and $\lambda-\lambda_j(z)$ linear factors of the generator's characteristic function. The $\lambda_j(1)$ are the system's relaxation rates. If the $z$-dependence is {\it not} independent of the parameter $s$, the assumption in (i) is falsified. One needs to start again at step (i) increasing, successively, from small values for $M$ and/or $m$ to larger values.

There are special cases of separable characteristic functions, see section~\ref{sec:SEBAndreev}, where different intervals of $s$-values reveal different dimensions and eigenvalues of the generator.
The total dimension and full spectrum is, then, obtained by combining the results from the different $s$-intervals.

\section{Conclusions}\label{sec:conclusions}
In this paper, we propose inverse counting statistics based on generalized factorial cumulants as a convenient and powerful tool to reconstruct characteristic features of a stochastic system from measured counting statistics of some of the system's transitions.
Such a method is particularly useful in cases in which very little is a priori known about the stochastic system under investigation.
As the only input information for the inverse counting-statistics procedure, we use a few experimentally determined numbers, namely the scaled generalized factorial cumulants in the long-time limit.
Despite the limited amount of input, the inverse counting-statistics procedure yields a remarkable extended amount of output.
First, we can determine a lower bound of $M$, the dimension of the stochastic system.
Second, we can find a lower bound of the counting power~$m$, which characterizes the coupling between stochastic system and detector.
Third, we can reconstruct the characteristic function $\chi(z,\lambda)$ of the generator $\mathbf{W}_z$, which is a polynomial of order $M$ in $\lambda$ and a polynomial of order $m$ in $z$.
From the zeros of the characteristic function, we can, then, determine the full $z$-dependence of the full spectrum of eigenvalues of  $\mathbf{W}_z$.
This is quite a remarkable result, since the long-time cumulants used as input depend only on one of the eigenvalues, $\lambda_\tx{max}$, determined around one value of $z$.

The use of generalized factorial cumulants instead of ordinary ones is crucial for our proposal.
While the evaluation of generalized factorial cumulants from a measured time trace of the detector signal does not introduce any extra complication as compared to the evaluation of ordinary cumulants, the benefit of having a free parameter $s$ in the definition of the generalized factorial cumulants is immense.
First, the outcome of the inverse counting statistics must be $s$-independent.
Therefore, an $s$-dependent outcome of the inverse counting statistics immediately indicates a wrong choice of $M$ or $m$.
Second, there are special cases of separable characteristic functions, for which the inverse counting statistics with ordinary cumulants would only reveal a part of the spectrum of eigenvalues of the generator, while the variation of $s$ makes it possible to access the full spectrum.

The proposed inverse counting-statistics procedure is quite general and, therefore, applicable to large variety of systems.
To illustrate the procedure we choose two examples from electronic transport in nanostructures: a single-level quantum dot in a Zeeman field and a single-electron box subjected to sequential and Andreev tunneling. 
For the latter case, the full dimension of the system's state space and the full spectrum of eigenvalues can only be revealed by varying the parameter $s$.

\ack
We acknowledge financial support from the Deutsche Forschungsgemeinschaft via grants KO 1987/5 and SFB1242.


\appendix

\section{Alternative expression for \texorpdfstring{$A_{l,\mu \nu}$}{\texttt{A\_\{l,mu nu\}}}} \label{sec:Apprandtel}
In the sum~(\ref{eq:As}) identical terms, related by permutation of the indices $(\alpha_1, \ldots, \alpha_\nu)$ may occur, e.g., $A_{\mu+2,\mu 2} = c_{s,0}c_{s,2}/2! + c_{s,1} c_{s,1} + c_{s,2}c_{s,0}/2! = c_{s,0}c_{s,2} + c_{s,1}^2$.
If one wants to combine them, one obtains the alternative representation
\begin{equation}
	A_{l,\mu \nu} = \sum_{{\beta_0 +\beta_1 +\ldots + \beta_\nu = \nu}\atop{\beta_1 +2\beta_2 \ldots + \nu \beta_\nu = l-\mu}} \frac{\nu!}{\beta_0! \cdot \beta_1! \cdots \beta_\nu!} \left( \frac{c_{s,0}}{0!}\right)^{\beta_0} \cdot \left( \frac{c_{s,1}}{1!}\right)^{\beta_1}\cdots \left( \frac{c_{s,\nu}}{\nu!}\right)^{\beta_\nu}\, .
\end{equation}

\section{Consistency check for \texorpdfstring{$\Delta=0.03 \, k_\tx{B}T$}{\texttt{Delta = 0.03 kBT}}}\label{sec:deltaapprox0}
\begin{figure*}[t]
\begin{center}
{\includegraphics[scale=1.00]{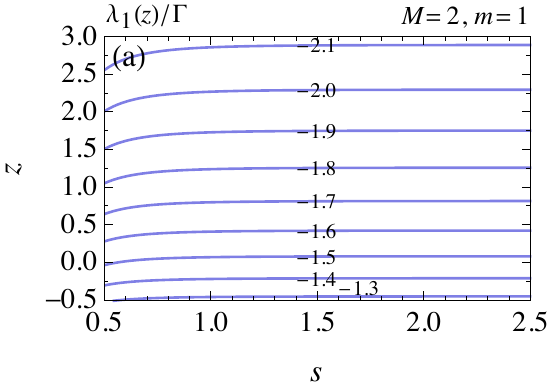}}
{\includegraphics[scale=1.00]{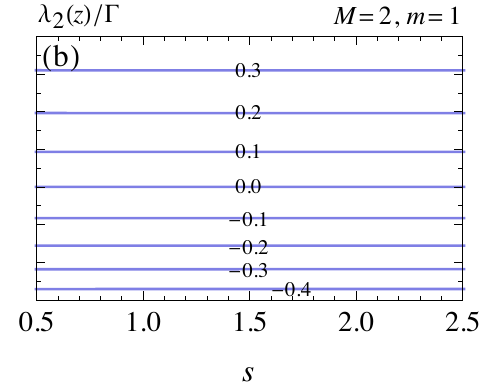}}
\caption{
	Consistency check of the number of system states~$M$ and the counting power~$m$ for the single-level quantum dot $\varepsilon=- k_\tx{B}T$ in an external magnetic field $\Delta=0.03\,k_\tx{B}T$ for time $\Gm t=100$. Assuming $M=2$ and $m=1$, contour lines of the resulting two eigenvalues $\lambda_1(z) / \, \Gm$ and $ \lambda_2(z) / \, \Gm$ are depicted.}
\label{fig10}
\end{center}
\end{figure*}

For completeness, we check how close the Zeeman field must be tuned to zero in order to observe the discussed behavior of section~\ref{sec:outCBlong}. We find that for $\Delta\lesssim 0.03\, k_\tx{B}T$, the eigenvalues obtained for $M=2,m=1$ are still almost $s$-independent (compare figure~\ref{fig10} to figure~\ref{fig5}), with slight deviations appearing for very negative $s$.

\section{Consistency check for \texorpdfstring{$n_\tx{G}=0.001$}{\texttt{nG = 0.001}}}\label{sec:ngapprox0}
\begin{figure*}
\begin{center}
{\includegraphics[scale=1.00]{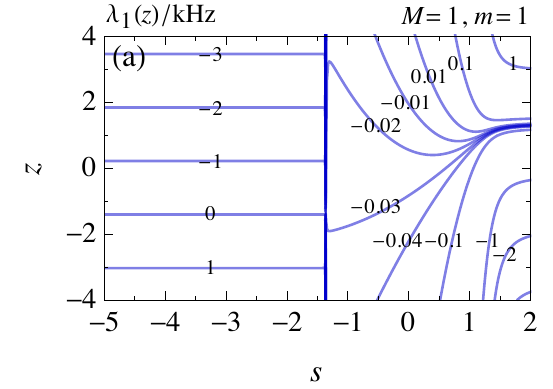}}
{\includegraphics[scale=1.00]{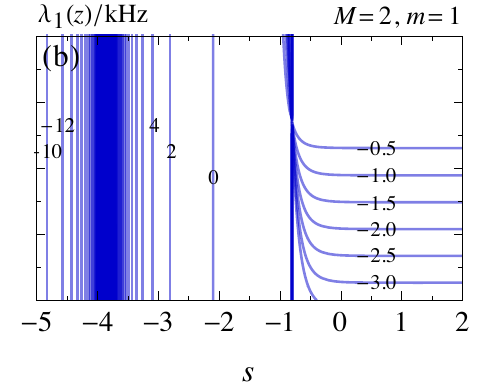}}
{\includegraphics[scale=1.00]{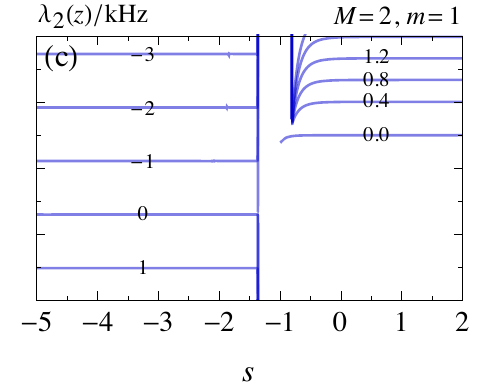}}
\caption{Consistency check of the number of system states~$M$ and the counting power~$m$ for the single-electron box with gate charge $n_\tx{G}=0.001$ for time $t=20\, \tx{s}$. Assuming different values for $M$ and $m$, contour lines of the resulting $j=1,\dots,M$ eigenvalues $ \lambda_j(z) / \, \tx{kHz}$ are depicted. The eigenvalue for $M=1$, $m=1$ in (a) and the two eigenvalues for $M=2$, $m=1$ in~(b), (c).}
\label{fig11}
\end{center}
\end{figure*}

For the single-electron box, we estimate that at least for $\Abs{n_\tx{G}}\lesssim 0.001$ the $n_\tx{G}=0$ case is already reached in good approximation for $s\lesssim -1.5$ or $s\gtrsim -0.5$ (compare figure~\ref{fig11} to figure~\ref{fig9}). For figure~\ref{fig11}, the Andreev-tunneling rates are approximated by $\Gm_\tx{A}^\pm \approx (615.11\pm 11.42) \, \tx{Hz} $~\cite{pekola_nonadiabatic_2008}. The sequential tunneling rates are estimated via an interpolation between the experimental values of~\onlinecite{maisi_real_2011}: $\Gm_\tx{u}^\pm \approx (12.00 \pm 0.03)\, \tx{Hz}$ and $\Gm_\tx{d}^\pm \approx (252.00 \pm 0.83)\, \tx{Hz}$.



\section*{References}
\begin{thebibliography}{65}

\bibitem{kolomeisky_molecular_2007} Kolomeisky~A~B and Fischer~M~E 2007
Molecular Motors: A Theorist's Perspective
\textit{Annu. Rev. Phys. Chem.} {\bf 58} 675

\bibitem{chemla_exact_2008} Chemla~Y~R, Moffitt~J~R and Bustamante~C 2008
Exact Solutions for Kinetic Models of Macromolecular Dynamics
\textit{J. Phys. Chem. B} {\bf 112} 6025

\bibitem{zhou_detecting_2011} Zhou~R, Kunzelmann~S, Webb~M~R and Ha~T 2011
Detecting Intramolecular Conformational Dynamics of Single Molecules in Short Distance Range with Subnanometer Sensitivity
\textit{Nano Lett.} {\bf 11} 5482

\bibitem{choi_single_2012} Choi~Y, Moody~I~S, Sims~P~C, Hunt~S~R, Corso~B~L, Perez~I, Weiss~G.~A and Collins~P~G 2012
Single-Molecule Lysozyme Dynamics Monitored by an Electronic Circuit
\textit{Science} {\bf 335} 319

\bibitem{chung_single_2012} Chung~H~S, McHale~K, Louis~J~M and Eaton~W~A 2012
Single-Molecule Fluorescence Experiments Determine Protein Folding Transition Path Times
\textit{Science} {\bf 335} 981

\bibitem{english_ever_2006} English~B~P, Min~W, van Oijen~A~M, Lee~K~T, Luo~G, Sun~H, Cherayil~B~J, Kou~S~C and Xie~X~S 2006
Ever-fluctuating single enzyme molecules: Michaelis-Menten equation revisited
\textit{Nature Chem. Biol.} {\bf 2} 87

\bibitem{moffitt_extracting_2014} Moffitt~R and Bustamante~C 2014
Extracting signal from noise: kinetic mechanisms from a Michaelis-Menten-like expression for enzymatic fluctuations
\textit{FEBS J.} {\bf 281} 498

\bibitem{cornish_survey_2007} Cornish~P~V and Ha~T 2007 
A Survey of Single-Molecule Techniques in Chemical Biology
\textit{ACS Chem. Biol.} {\bf 2} 53

\bibitem{kim_single_2013} Kim~H and Ha~T 2013
Single-molecule nanometry for biological physics
\textit{Rep. Prog. Phys.} {\bf 76} 016601


\bibitem{levitov_electron_1996} Levitov~L~S, Lee~H and Lesovik~G~B 1996
Electron counting statistics and coherent states of electric current
\textit{J. Math. Phys.} {\bf 37} 4845

\bibitem{bagrets_full_2003} Bagrets~D~A and Nazarov~Yu~V 2003
Full counting statistics of charge transfer in Coulomb blockade systems
\textit{Phys. Rev. B} {\bf 67} 085316


\bibitem{lindebaum_spin-induced_2009} Lindebaum~S, Urban~D and K{\"o}nig~J 2009
Spin-induced charge correlations in transport through interacting quantum dots with ferromagnetic leads
\textit{Phys. Rev. B} {\bf 79} 245303

\bibitem{schmidt_charge_2009} Schmidt~T~L and Komnik~A 2009
Charge transfer statistics of a molecular quantum dot with a vibrational degree of freedom
\textit{Phys. Rev. B} {\bf 80} 041307(R)

\bibitem{schaller_counting_2010} Schaller~G, Kie{\ss}lich~G and Brandes~T 2010
Counting statistics in multistable systems
\textit{Phys. Rev. B} {\bf 81} 205305

\bibitem{belzig_full_2005} Belzig~W 2005
Full counting statistics of super-Poissonian shot noise in multilevel quantum dots
\textit{Phys. Rev. B} {\bf 71} 161301(R)

\bibitem{urban_coulomb-interaction_2008} Urban~D, K{\"o}nig~J and Fazio~R 2008
Coulomb-interaction effects in full counting statistics of a quantum-dot Aharonov-Bohm interferometer
\textit{Phys. Rev. B} {\bf 78} 075318

\bibitem{belzig_full_counting_2001}
Belzig~W and Nazarov~Yu~V 2001
Full Counting Statistics of Electron Transfer between Superconductors
\textit{Phys. Rev. Lett.} {\bf 87} 197006

\bibitem{boerlin_full_2002}
B\"orlin~J, Belzig~W and Bruder~C 2002
Full Counting Statistics of a Superconducting Beam Splitter
\textit{Phys. Rev. Lett.} {\bf 88} 197001

\bibitem{cuevas_full_2003}
Cuevas~J~C and Belzig~W 2003
Full Counting Statistics of Multiple Andreev Reflections
\textit{Phys. Rev. Lett.} {\bf 91} 187001

\bibitem{johansson_full_2003}
Johansson~G, Samuelsson~P and Ingerman~\AA{} 2003
Full Counting Statistics of Multiple Andreev Reflection
\textit{Phys. Rev. Lett.} {\bf 91} 187002

\bibitem{pilgram_noise_2005}
Pilgram~S and Samuelsson~P 2005
Noise and Full Counting Statistics of Incoherent Multiple Andreev Reflection
\textit{Phys. Rev. Lett.} {\bf 94} 086806.

\bibitem{morten_full_2008}
Morten~J~P, Huertas-Hernando~D, Belzig~W and Brataas~A 2008
Full counting statistics of crossed Andreev reflection
\textit{Phys. Rev. B} {\bf 78} 224515

\bibitem{braggio_superconducting_2011} Braggio~A, Governale~M, Pala~M~G and K{\"o}nig~J 2011
Superconducting proximity effect in interacting quantum dots revealed by shot noise
\textit{Solid State Commun.} {\bf 151} 155

\bibitem{poeltl_feedback_2011} P{\"o}ltl~C, Emary~C and Brandes~T 2011
Feedback stabilization of pure states in quantum transport
\textit{Phys. Rev. B} {\bf 84} 085302

\bibitem{daryanoosh_stochastic_2016} Daryanoosh~S, Wiseman~H~M and Brandes~T 2016
Stochastic feedback control of quantum transport to realize a dynamical ensemble of two nonorthogonal pure states
\textit{Phys. Rev. B} {\bf 93} 085127

\bibitem{wagner_squeezing_2016} Wagner~T, Strasberg~P, Bayer~J~C, Rugeramigabo~E~P, Brandes~T and Haug~R~J 2016
Strong suppression of shot noise in a feedback-controlled single-electron transistor,
\textit{Nature Nanotechnol.} advance online publication


\bibitem{gustavsson_counting_2005} Gustavsson~S, Leturcq~R, Simovi\v{c}~B, Schleser~R, Ihn~T, Studerus~P, Ensslin~K, Driscoll~D~C and Gossard~A~C 2006
Counting Statistics of Single Electron Transport in a Quantum Dot
\textit{Phys. Rev. Lett.} {\bf 96} 076605

\bibitem{fujisawa_bidirectional_2006} Fujisawa~T, Hayashi~T, Tomita~R and Hirayama~Y 2006
Bidirectional Counting of Single Electrons
\textit{Science} {\bf 312} 1634

\bibitem{gustavsson_measurements_2007} Gustavsson~S, Leturcq~R, Ihn~T, Ensslin~K, Reinwald~M and Wegscheider~W 2007
Measurements of higher-order noise correlations in a quantum dot with a finite bandwidth detector
\textit{Phys. Rev. B} {\bf 75} 075314

\bibitem{fricke_bimodal_2007}
Fricke~C, Hohls~F, Wegscheider~W and Haug~R~J 2007
Bimodal counting statistics in single-electron tunneling through a quantum dot
\textit{Phys. Rev. B} {\bf 76} 155307

\bibitem{flindt_universal_2009} Flindt~C, Fricke~C, Hohls~F, Novotn{\'y}~T, Neto{\v c}n{\'y}~K, Brandes~T and Haug~R~J 2009
Universal oscillations in counting statistics
\textit{PNAS} {\bf 106} 10116

\bibitem{gustavsson_electron_2009} Gustavsson~S, Leturcq~M, Studer~R, Shorubalko~I, Ihn~T, Ensslin~K, Driscoll~D~C and Gossard~A~C 2009
Electron counting in quantum dots
\textit{Surf. Sci. Rep.} {\bf 64} 191

\bibitem{fricke_high_2010} Fricke~C, Hohls~F, Flindt~C and Haug~R~J 2010
High cumulants in the counting statistics measured for a quantum dot
\textit{Physica E} {\bf 42} 848

\bibitem{fricke_high-order_2010} Fricke~C, Hohls~F, Sethubalasubramanian~N, Fricke~L and Haug~R~J 2010
High-order cumulants in the counting statistics of asymmetric quantum dots
\textit{Appl. Phys. Lett.} {\bf 96} 202103

\bibitem{komijani_counting_2013} Komijani~Y, Choi~T, Nichele~F, Ensslin~K, Ihn~T, Reuter~D and Wieck~A~D 2013
Counting statistics of hole transfer in a $p$-type GaAs quantum dot with dense excitation spectrum
\textit{Phys. Rev. B} {\bf 88} 035417


\bibitem{martinis_metrological_1994}
Martinis~J~M, Nahum~M and Jensen~H~D 1994
Metrological accuracy of the electron pump
\textit{Phys. Rev. Lett.} {\bf 72} 904

\bibitem{dresselhaus_measurement_1994}
Dresselhaus~P~D, Ji~L, Han~S, Lukens~J~E and Likharev~K~K 1994
Measurement of single electron lifetimes in a multijunction trap
\textit{Phys. Rev. Lett.} {\bf 72} 3226

\bibitem{lotkhov_storage_1999}
Lotkhov~S~V, Zangerle~H, Zorin~A~B and Niemeyer~J 1999
Storage capabilities of a four-junction single-electron trap with an on-chip resistor
\textit{Appl. Phys. Lett.} {\bf 75} 2665

\bibitem{lu_real_2003}
Lu~W, Ji~Z, Pfeiffer~L, West~K~W and Rimberg~A~J 2003
Real-time detection of electron tunnelling in a quantum dot 
\textit{Nature (London)} {\bf 423} 422

\bibitem{bylander_current_2005}
Bylander~J, Duty~T and Delsing~P 2005
Current measurement by real-time counting of single electrons
\textit{Nature (London)} {\bf 434} 361



\bibitem{kurzmann_optical_2016} Kurzmann~A, Merkel~B, Labud~P~A, Ludwig~A, Wieck~A~D, Lorke~A and Geller~M 2016
Optical Blocking of Electron Tunneling into a Single Self-Assembled Quantum Dot
\textit{Phys. Rev. Lett.} {\bf 117} 017401

\bibitem{dasenbrook_dynamical_2016} Dasenbrook~D and Flindt~C 2016
Dynamical Scheme for Interferometric Measurements of Full Counting Statistics
\textit{Phys. Rev. Lett.} {\bf 117} 146801


\bibitem{bruderer_inverse_2014} Bruderer~M, Contreras-Pulido~L~D, Thaller~M, Sironi~L, Obreschkow~D and Plenio~M~B 2014
Inverse counting statistics for stochastic and open quantum systems: the characteristic polynomial approach
\textit{New J. Phys.} {\bf 16} 033030

\bibitem{splettstoesser_charge_2010}
Splettstoesser~J, Governale~M, K\"onig~J and B\"uttiker~M 2010
Charge and spin dynamics in interacting quantum dots
\textit{Phys. Rev. B} {\bf 81} 165318

\bibitem{pulido_time_2012}
Contreras-Pulido~L~D, Splettstoesser~J, Governale~M, K\"onig~J and B\"uttiker~M 2012
Time scales in the dynamics of an interacting quantum dot
\textit{Phys. Rev. B} {\bf 85} 075301

\bibitem{schulenborg_detection_2014}
Schulenborg~J, Splettstoesser~J, Governale~M and Contreras-Pulido~L~D 2014
Detection of the relaxation rates of an interacting quantum dot by a capacitively coupled sensor dot
\textit{Phys. Rev. B} {\bf 89} 195305

\bibitem{schulenborg_parity_2016}
Schulenborg~J, Saptsov~R~B, Haupt~F, Splettstoesser~J and Wegewijs~M~R 2016
Fermion-parity duality and energy relaxation in interacting open systems
\textit{Phys. Rev. B} {\bf 93} 081411(R)

\bibitem{riwar_readout_2016} Riwar~R-P, Roche~B, Jehl~X and Splettstoesser~J 2016
Readout of relaxation rates by nonadiabatic pumping spectroscopy
\textit{Phys. Rev. B} {\bf 93} 235401

\bibitem{vanherck_relaxation_2016} Vanherck~J, Schulenborg~J, Saptsov~R~B, Splettstoesser~J and Wegewijs~M~R 2016
Relaxation of quantum dots in a magnetic field at finite bias -Charge, spin, and heat currents 
\textit{Phys. Status Solidi B} (https://doi.org/10.1002/pssb.201600614)


\bibitem{feve_on-demand_2007} F{\`e}ve~G, Mah{\'e}~A, Berroir~J-M, Kontos~T, Pla{\c c}ais~B, Glattli~D~C, Cavanna~A, Etienne~B and Jin~Y 2007
An On-Demand Coherent Single-Electron Source
\textit{Science} {\bf 316} 1169

\bibitem{mahe_current_2010} Mah\'e~A, Parmentier~F~D, Bocquillon~E, Berroir~J-M, Glattli~D~C, Kontos~T, Pla\ifmmode \mbox{\c{c}}\else \c{c}\fi{}ais~B, F\`eve~G, Cavanna~A and Jin~Y 2010
 Current correlations of an on-demand single-electron emitter
\textit{Phys. Rev. B} {\bf 82} 201309(R)

\bibitem{beckel_asymmetry_2014} Beckel~A, Kurzmann~A, Geller~M, Ludwig~A, Wieck~A~D, K{\"o}nig~J and Lorke~A 2014
Asymmetry of charge relaxation times in quantum dots: The influence of degeneracy
\textit{Europhys. Lett.} {\bf 106} 47002

\bibitem{hofmann_measuring_2016} Hofmann~A, Maisi~V~F, Gold~C, Kr\"ahenmann~T, R\"ossler~C, Basset~J, M\"arki~P, Reichl~C, Wegscheider~W, Ensslin~K and Ihn~T 2016
Measuring the Degeneracy of Discrete Energy Levels Using a $\mathrm{GaAs}/\mathrm{AlGaAs}$ Quantum Dot
\textit{Phys. Rev. Lett.} {\bf 117} 206803

\bibitem{kurzmann_electron_2017} Kurzmann~A, Merkel~B, Marquardt~B, Beckel~A, Ludwig~A, Wieck~A~D, Lorke~A and Geller~M 2017
Electron dynamics in transport and optical measurements of self-assembled quantum dots \textit{Phys. Status Solidi B} (https://doi.org/10.1002/pssb.201600625)


\bibitem{stegmann_detection_2015} Stegmann~P, Sothmann~B, Hucht~A and K\"onig~J 2015
Detection of interactions via generalized factorial cumulants in systems in and out of equilibrium
\textit{Phys. Rev. B} {\bf 92} 155413

\bibitem{stegmann_short_2016} Stegmann~P and K\"onig~J 2016
Short-time counting statistics of charge transfer in Coulomb-blockade systems
\textit{Phys. Rev. B} {\bf 94} 125433

\bibitem{braggio_full_2006} Braggio~A, K{\"o}nig~J and Fazio~R 2006
Full Counting Statistics in Strongly Interacting Systems: Non-Markovian Effects
\textit{Phys. Rev. Lett.} {\bf 96} 026805

\bibitem{flindt_counting_2008} Flindt~C, Novotn{\'y}~T, Braggio~A, Sassetti~M and Jauho~A-P 2008
Counting Statistics of Non-Markovian Quantum Stochastic Processes
\textit{Phys. Rev. Lett.} {\bf 100} 150601

\bibitem{flindt_counting_2010} Flindt~C, Novotn{\'y}~T, Braggio~A and Jauho~A-P 2010 
Counting statistics of transport through Coulomb blockade nanostructures: High-order cumulants and non-Markovian effects
\textit{Phys. Rev. B} {\bf 82} 155407

\bibitem{sothmann_nonequilibrium_2010} Sothmann~B and K\"onig~J 2010
Nonequilibrium current and noise in inelastic tunneling through a magnetic atom
\textit{New J. Phys.} {\bf 12} 083028

\bibitem{kambly_factorial_2011} Kambly~D, Flindt~C and B{\"u}ttiker~M 2011
Factorial cumulants reveal interactions in counting statistics
\textit{Phys. Rev. B} {\bf 83} 075432

\bibitem{kambly_time-dependent_2013} Kambly~D and Flindt~C 2013
Time-dependent factorial cumulants in interacting nano-scale systems
\textit{J. Comput. Electron.} {\bf 12} 331

\bibitem{droste_finite_2016} Droste~S and Governale~M 2016
Finite-time full counting statistics and factorial cumulants for transport through a quantum dot with normal and superconducting leads
\textit{J. Phys.: Condens. Matter} {\bf 28} 145302


\bibitem{saira_environmentally_2010}
Saira~O-P, M\"ott\"onen~M, Maisi~V~F and Pekola~J~P 2010
Environmentally activated tunneling events in a hybrid single-electron box
\textit{Phys. Rev. B} {\bf 82} 155443

\bibitem{maisi_real_2011}
Maisi~V~F, Saira~O-P, Pashkin~Yu~A, Tsai~J~S, Averin~D~V and Pekola~J~P 2011
Real-Time Observation of Discrete Andreev Tunneling Events
\textit{Phys. Rev. Lett.} {\bf 106} 217003

\bibitem{maisi_full_2014}
Maisi~V~F, Kambly~D, Flindt~C and Pekola~J~P 2014
Full Counting Statistics of Andreev Tunneling
\textit{Phys. Rev. Lett.} {\bf 112} 036801

\bibitem{stegmann_violation_2016} Stegmann~P and K\"onig~J 2016
Violation of detailed balance for charge-transfer statistics in Coulomb-blockade systems
\textit{Phys. Status Solidi B} (https://doi.org/10.1002/pssb.201600507)

\bibitem{pekola_environment_2010}
Pekola~J~P, Maisi~V~F, Kafanov~S, Chekurov~N, Kemppinen~A, Pashkin~Yu~A, Saira~O-P, M\"ott\"onen~M and Tsai~J~S 2010
Environment-Assisted Tunneling as an Origin of the Dynes Density of States
\textit{Phys. Rev. Lett.} {\bf 105} 026803

\bibitem{pekola_nonadiabatic_2008} Averin D~V and Pekola~J~P 2008
Nonadiabatic Charge Pumping in a Hybrid Single-Electron Transistor
\textit{Phys. Rev. Lett.} {\bf 101} 066801

\end{thebibliography}
\end{document}